\documentclass[lettersize,journal]{IEEEtran}
\usepackage{cite}
\usepackage{bm}
\usepackage{amsmath,amsfonts}
\usepackage[noend]{algorithmic}
\usepackage{algorithm}
\usepackage{array}
\usepackage{subfig}
\usepackage{textcomp}
\usepackage{stfloats}
\usepackage{url}
\usepackage{verbatim}
\usepackage{graphicx}
\usepackage[font=small]{caption}  
\usepackage{booktabs} 
\usepackage{makecell}
\usepackage{multirow}
\usepackage{stfloats}
\usepackage{hyperref}
\usepackage{amssymb}

\usepackage{threeparttable}

\begin{document}

\title{SPAT: A Semantic Port-Aware Adaptive-Rate Transmission Protocol for Semantic Communication}

\author{
Yunhao Wang,~\IEEEmembership{Graduate Student Member,~IEEE,}
Shuai Ma,~\IEEEmembership{Member,~IEEE,} \\
Bin Shen,~\IEEEmembership{Graduate Student Member,~IEEE,}
Shouhan Shi,~\IEEEmembership{Graduate Student Member,~IEEE,} \\
Youlong Wu,~\IEEEmembership{Member,~IEEE,}
Guangming Shi,~\IEEEmembership{Fellow,~IEEE,}
and Xiang Cheng,~\IEEEmembership{Fellow,~IEEE} \\

\thanks{This work was supported in part by the National Science and Technology Major Project-Mobile Information Networks under Grant No.2024ZD1300700, and in part by the Natural Science Foundation of China No.62293483. \textit{(Corresponding Author: Guangming Shi.)}}
\thanks{Yunhao Wang is with the School of Electronic and Computer Engineering, Peking University, Shenzhen 518055, China (e-mail: yunhaowang@stu.pku.edu.cn).}
\thanks{Shuai Ma is with the Department of Networked Intelligence, Peng Cheng Laboratory, Shenzhen 518066, China (e-mail: mash01@pcl.ac.cn).}
\thanks{Bin Shen  and  Shouhan Shi are with the School of Information and Control Engineering, China University of Mining and Technology, Xuzhou, 221116, China (e-mail: \{binshen, shishouhan\}@cumt.edu.cn).}
\thanks{Youlong Wu is with the School of Information Science and Technology, ShanghaiTech University, Shanghai 201210, China (e-mail:wuyl1@shanghaitech.edu.cn).}
\thanks{Guangming Shi is with the Department of Networked Intelligence, Peng Cheng Laboratory, Shenzhen, 518066, China, and also with the School of Artificial Intelligence, Xidian University, Xi’an, Shaanxi 710071, China (e-mail: gmshi@pcl.ac.cn).}
\thanks{Xiang Cheng is with the State Key Laboratory of Photonics and Communications, School of Electronics, Peking University, Beijing 100871, China (e-mail: xiangcheng@pku.edu.cn).}
\thanks{The code will be released on \url{https://github.com/WYHxuebi/SPAT}.}
}

\maketitle

\begin{abstract}
With the evolution of 6G, semantic communication has emerged as a promising paradigm by prioritizing the delivery of task-relevant meaning over strict bit-level correctness. However, existing transport mechanisms still rely on explicit port headers and bit-level validation, making them vulnerable to header corruption and the resulting packet loss. To address this issue, this paper proposes a Semantic Port-Aware Adaptive-Rate Transmission Protocol (SPAT) for semantic communication. The proposed framework jointly embeds source and destination port information into semantic representations, thereby reducing dependence on explicit port headers while enabling robust port-aware transmission. Furthermore, a differentiated semantic processing mechanism is developed for uplink and downlink scenarios, where port identification is introduced for uplink service recognition and destination-aware conditional gating is designed for downlink selective decoding. In addition, an adaptive-rate controller is incorporated to dynamically adjust the number of transmitted semantic channels according to channel conditions and feature importance, thereby improving both robustness and transmission efficiency. Experimental results on the AFHQ and ImageNet-10 datasets, together with real-world experimental measurements, demonstrate that SPAT consistently outperforms TCP, UDP, and SITP in reconstruction quality across different SNRs while maintaining low-latency transmission.
\end{abstract}



\section{Introduction}

\IEEEPARstart{W}{ith} the evolution toward sixth-generation (6G), emerging applications such as the industrial Internet of Things \cite{chen2026role}, edge intelligence \cite{yan2026flexible}, and digital twins \cite{li2025generative} are driving wireless systems from connectivity-oriented architectures toward intelligence-native paradigms. Recent studies indicate that future networks require latency around 0.1 ms, while reliability is expected to approach $10^{-7}$ or even higher levels \cite{alliance2024itu, tao20236g}. However, conventional communication remain centered on bit-level accurate transmission, which limits the ability to simultaneously achieve ultra-low latency and high resource efficiency. In contrast, semantic communication (SemCom) shifts the design from bit-level correctness to semantic effectiveness, emerging as a promising paradigm for intelligent communication in future 6G networks \cite{liu2024survey, guo2025semantic}.

\begin{figure}[t]
    \centering
    \includegraphics[width=0.49\textwidth]{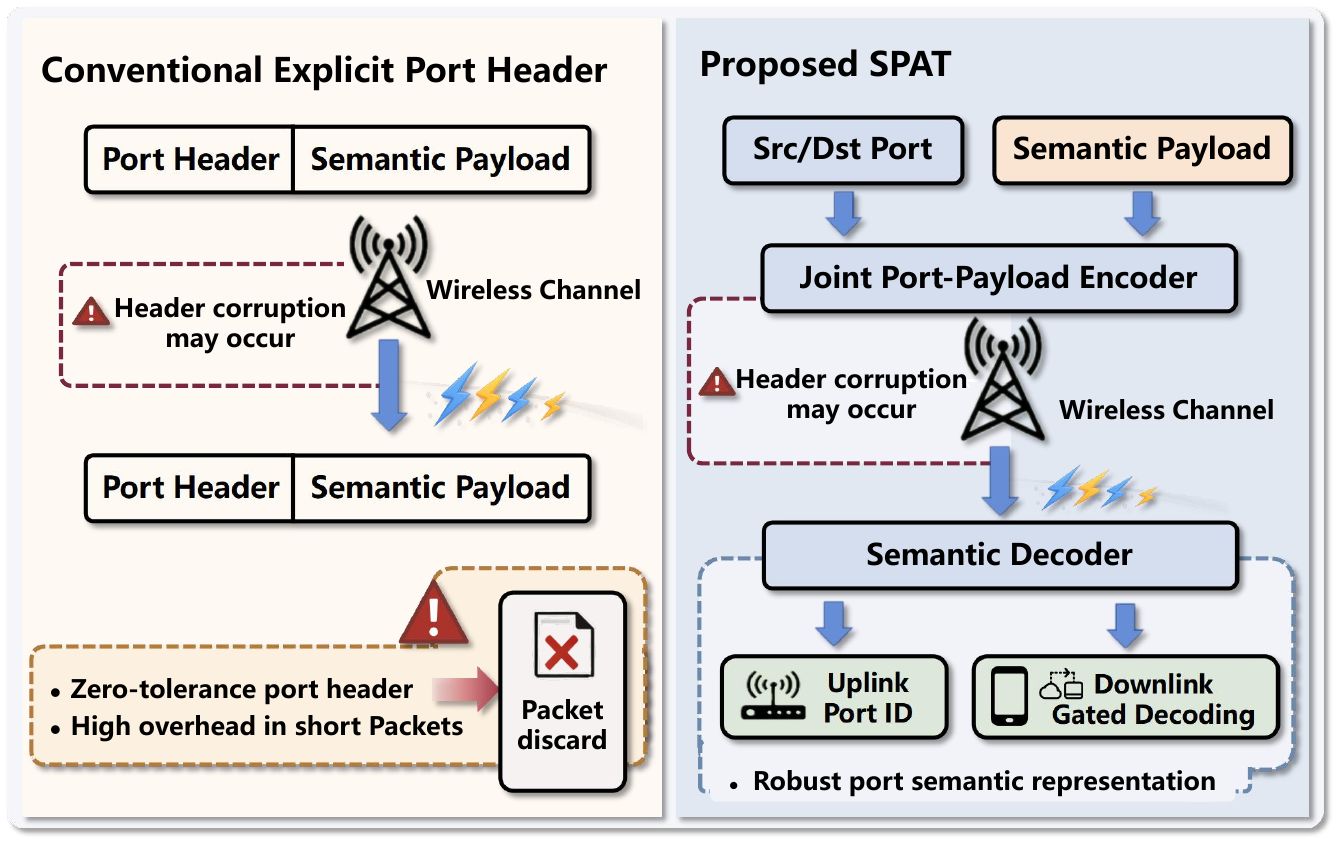}
    \caption{Comparison between conventional explicit port-header transmission and the proposed SPAT \textbf{(Ours)}.}
    \label{fig:ProtocolComparison}
\end{figure}

SemCom represents a paradigm shift toward the effective delivery of task-relevant semantic information \cite{getu2025semantic}. Unlike conventional communication rooted in Shannon’s information theory \cite{shannon1948mathematical}, which mainly focus on reliable bit transmission, SemCom emphasizes the meaning and task relevance of the conveyed information \cite{wang2024swin}. With the rapid advancement of deep learning, data-driven SemCom have made it possible to learn robust semantic extraction, representation, and recovery: \textit{\textbf{even when semantic features are partially corrupted by noise, the receiver can still recover semantically consistent information for downstream task execution \cite{wang2025noc4sc, ma2023task}.}}

Recent SemCom research has attracted considerable attention, particularly in end-to-end representation learning, which has been extended to various applications, including image \cite{lyu2024semantic, bian2023deepjscc}, text \cite{zhu2025stochastic}, video \cite{li2026goal}, point cloud \cite{yang2026channel}, and multimodal tasks \cite{fang2026cross}. However, most existing works remain confined to physical-layer abstractions, primarily addressing wireless channel impairments while overlooking packetization and multi-layer protocol mechanisms in practical systems. Although several studies have begun to investigate cross-layer modeling for digital SemCom \cite{teng2025conquering, wang2025sitp}, \textit{\textbf{a systematic framework for semantic representation and transmission of transport-layer port information, particularly source and destination port information, remains largely unexplored.}}

From a cross-layer perspective, conventional protocols such as TCP \cite{tian2005tcp} enhance reliability through retransmission, whereas UDP \cite{gu2007udt} reduces transmission delay by adopting a lightweight mechanism. Recently, transport schemes, such as the semantic information transport protocol (SITP) \cite{wang2025sitp}, have been introduced to better match the transmission requirements of semantic information. Nevertheless, existing transport protocols remain payload-centric, treating metadata such as port information through explicit headers that require bit-level correctness. \textit{\textbf{As a result, corruption of critical header bits often leads to packet discard, which is fundamentally inconsistent with the error-tolerant nature of semantic payloads.}}

\captionsetup[table]{justification=centering, labelsep=space, textfont=sc} 
\begin{table*}[t]
    \centering
    \renewcommand\arraystretch{2.2}
    \setlength{\tabcolsep}{5pt}
    \caption{ \\ Comparison of Conventional Protocols and the Proposed SPAT Framework \label{tab1}}
    \begin{tabular}{ccccccc}
        \hline
        \hline
        \textbf{Protocol} & \textbf{Paradigm} & \textbf{Handshake \& Retrans.} & \textbf{Port Rep.} & \makecell{\textbf{Short-Packet} \\ \textbf{Overhead}} & \textbf{Packet Validation Scope} & \makecell{\textbf{Uplink/Downlink} \\ \textbf{Differentiation}} \\
        \hline
        TCP \cite{tian2005tcp} & Bit-oriented & \makecell{3-way handshake \\ and ACK retrans.} & Explicit & High & Header + Payload & No \\
        UDP \cite{gu2007udt} & Bit-oriented & \makecell{No handshake \\ or retrans.} & Explicit & High & Header + Payload & No  \\
        UDP-Lite \cite{larzon1999udp} & Bit-oriented & \makecell{No handshake \\ or retrans.} & Explicit & High & Header + Partial payload & No \\
        PR-SCTP \cite{wang2005applying} & Bit-oriented & \makecell{4-way handshake \\ and partial retrans.} & Explicit & High & Header + Payload & No \\
        SITP \cite{wang2025sitp} & Semantic-oriented & \makecell{No handshake \\ or retrans.} & Explicit & High & Header-only & No \\
        \textbf{SPAT (Ours)} & \textbf{Semantic-oriented} & \makecell{\textbf{No handshake} \\ \textbf{or retrans.}} & \makecell{\textbf{Implicit semantic} \\ \textbf{embedding}} & \textbf{Low} & \textbf{Semantic port decoding} & \textbf{Yes} \\
        \hline
        \hline
    \end{tabular}
    \begin{tablenotes}
        \footnotesize
        \item[*] \textit{Note:} \textbf{Retrans.} denotes retransmissions, and \textbf{Rep.} denotes representation.
    \end{tablenotes}
\end{table*}

Explicit header mechanisms introduce additional challenges in short-packet communications \cite{vu2025short}. In applications such as industrial IoT and edge sensing, the small payload size makes fixed-format transport headers a substantial source of overhead, thereby reducing semantic transmission efficiency \cite{krekovic2025reducing}. Moreover, port information serves distinct functions in uplink and downlink transmissions: the uplink mainly requires service identification, whereas the downlink focuses on selective reception and decoding at the target port \cite{yu2024fdlora}. However, traditional explicit-header designs generally adopt a uniform treatment, which makes it difficult to accommodate these distinct semantic requirements.

Overall, existing research still faces several key challenges:
\begin{itemize}
\item{The bit-level accuracy required for transport-layer headers is inherently inconsistent with the error-tolerant nature of semantic payloads.}
\item{In short-packet scenario, explicit headers introduce considerable overhead, limiting transmission efficiency.}
\item{Uplink and downlink transmissions impose distinct semantic requirements on port information, while dedicated design remain underexplored.}
\end{itemize}

\subsection{Contributions}

To the best of our knowledge, existing studies have rarely considered semantic representations for transport-layer port information. In this work, we focus on source and destination port numbers, which are essential for service identification and delivery. The main contributions are summarized as follows:

\begin{itemize}

\item{\textbf{Semantic Port-Aware Adaptive-Rate Transmission Protocol (SPAT):} We propose a joint port-payload embedding framework for SemCom, where the source and destination port are mapped into a high-dimensional semantic space and jointly encoded with the semantic payload to enable implicit port transmission. Compared with explicit-header designs, the proposed framework reduces dependence on bit-level correct headers while preserving service identifiability.}

\item{\textbf{Port-aware transmission for uplink and downlink:} We propose a differentiated port-aware semantic transmission mechanism for uplink and downlink. For the uplink, a port identification module is designed to enable the base station to recognize service sources and destination port information. For the downlink, a port-aware gated decoder is developed to selectively decode semantic content associated with the target port.}

\item{\textbf{Adaptive importance-based rate transmission:} We develop an adaptive rate transmission mechanism that dynamically adjusts the number of transmitted semantic channels according to channel conditions and feature importance, thereby balancing transmission efficiency and reconstruction quality under varying channel conditions.}

\item{\textbf{Extensive experiments} together with real-world testbed measurements demonstrate that the proposed SPAT framework achieves lower latency than TCP, UDP, and SITP while providing superior reconstruction performance. The adaptive-rate mechanism further improves transmission robustness and efficiency through flexible channel allocation under varying channel conditions.}

\end{itemize}

\subsection{Organization of This Article}

Section III presents the proposed port-aware SemCom system. Section IV presents the uplink and downlink semantic transmission framework, including port embedding, port identification, and conditional gating for selective decoding. Section V develops the adaptive-rate control mechanism. Section VI provides the simulation results and performance analysis. Section VII provides the real-world experimental results. Finally, Section VIII concludes this paper.

\begin{figure*}[t]
\centering
\includegraphics[width=0.95\textwidth]{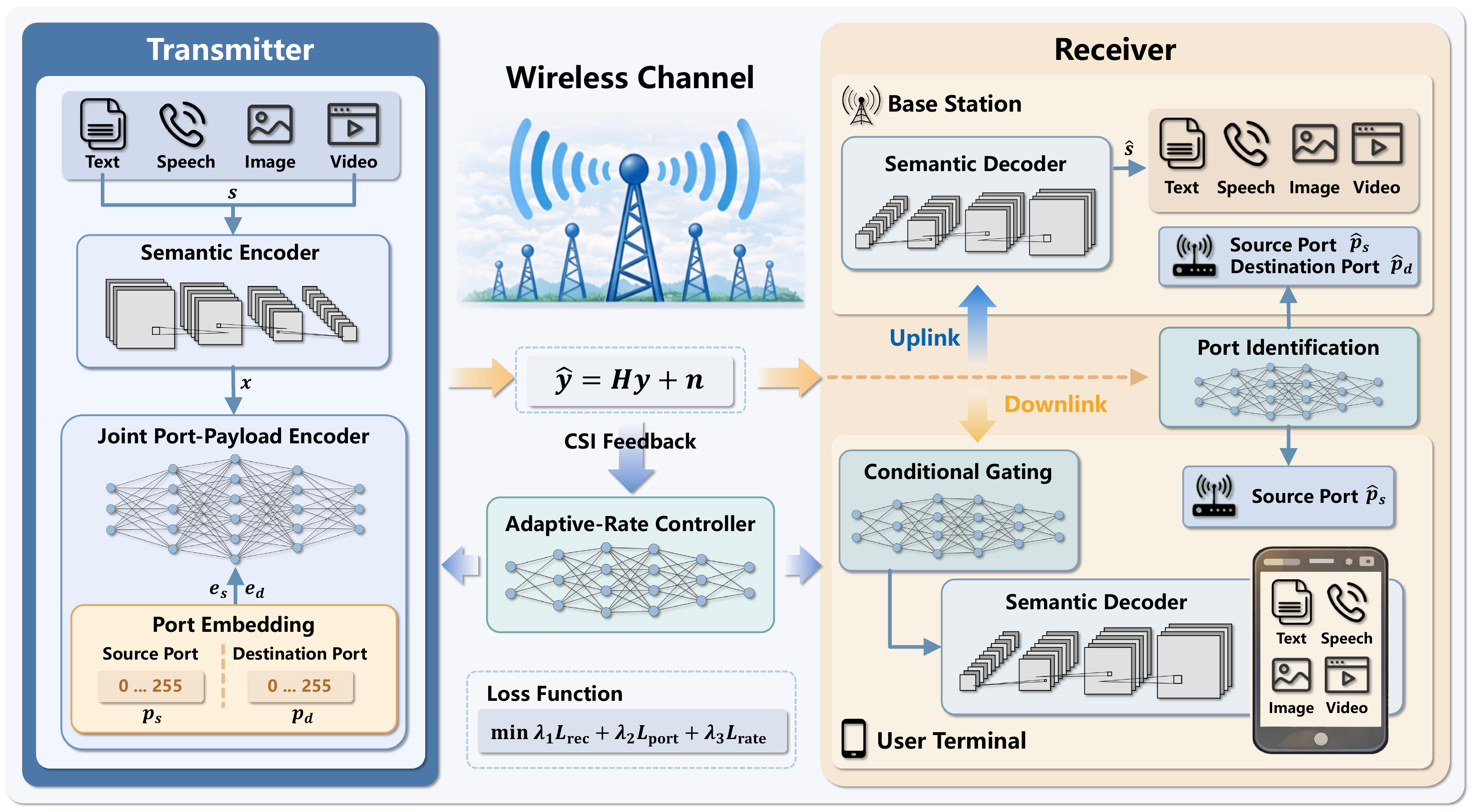}
\caption{The overall architecture of the SPAT-enabled SemCom framework. The proposed SPAT framework embeds port information into semantic features via joint port–payload encoding, enabling port-aware semantic delivery with port identification and conditional gating for differentiated uplink and downlink decoding at the receiver.}
\label{fig:Architecture}
\end{figure*}

\section{Related Work}

\subsection{Transport-Layer Protocols}

Current transport-layer protocols are primarily built upon classical schemes such as TCP and UDP, both of which are designed for bit-level reliable packet delivery \cite{haryono2025comparative}. TCP ensures high reliability through connection establishment, acknowledgement, and retransmission, but incurs substantial latency and protocol overhead \cite{tian2005tcp}. In contrast, UDP adopts a connectionless and lightweight design without retransmission, offering lower latency but weaker reliability, since packets are typically discarded once checksum verification fails \cite{gu2007udt}. Both protocols rely on explicit transport-layer headers and strict packet integrity, such that errors in either header or payload bits directly result in packet loss. While this design is effective for conventional bit-oriented communications, it is less suitable for SemCom systems, where semantic features may remain recoverable under partial distortion.

Beyond TCP and UDP, several transport-layer variants have been introduced to accommodate diverse application requirements \cite{10.1145/2333112.2333113, wang2005applying, larzon1999udp}. SCTP offers message-oriented and multistream transport with congestion and reliability control \cite{10.1145/2333112.2333113}, while PR-SCTP relaxes full reliability through partial retransmission of application-defined chunks \cite{wang2005applying}. UDP-Lite reduces checksum coverage over part of the payload, allowing limited payload corruption while still preserving header correctness \cite{larzon1999udp}. However, such protocols still rely on explicit bit-level headers, which must be received correctly for successful packet interpretation and delivery.

Recently, semantic-oriented transport protocols have begun to emerge. In particular, SITP enhances semantic usability by prioritizing header verification while allowing noisy semantic payloads to remain deliverable \cite{wang2025sitp}. However, SITP still relies on explicit transport-layer headers to carry source and destination port information, and packet delivery remains dependent on bit-level header correctness. Consequently, corruption in critical header bits may still lead to service identification failure or packet discard. Overall, both conventional and semantic-aware transport protocols still represent and protect port information in an explicit bit-level manner. To fill this gap, the proposed SPAT framework embeds port information into the semantic payload through semantic representation, thereby reducing reliance on bit-level correct explicit port headers.

\subsection{Semantic Communication}

Recent advances in SemCom have mainly focused on end-to-end representation learning under complex wireless channels \cite{peng2025robust, jin2025task, park2025transmit}. In \cite{peng2025robust}, DeepSC-RI employs a multi-scale vision transformer with dual-branch semantic extraction to enable robust semantic transmission. In \cite{jin2025task}, SCASR improves task-oriented SemCom by adaptively compressing and reconstructing semantic features under varying SNR, compression rate, and packet loss conditions. \cite{park2025transmit} proposes a task-adaptive SemCom framework for visual information, which selects task-relevant semantics according to the target task. However, most existing SemCom frameworks are developed over abstract physical-layer channels, focusing mainly on robustness to noise and fading while seldom considering packetization or packet discard in practical protocol stacks.

A few studies have extended SemCom beyond purely noisy-channel settings. The packet-loss-resilient video SemCom system MSTVSC was proposed in \cite{teng2025conquering}, where a MoE Swin Transformer and packet-loss recovery are employed to preserve reconstruction quality. However, most existing works still rely on simplified erasure models that directly remove lost packets and only evaluate their impact on semantic reconstruction. Overall, current SemCom research remains focused on end-to-end physical-layer channel models, with limited consideration of packet loss caused by header bit errors. More importantly, the semantic representation and integration of source and destination port information into the transmission process have rarely been explored.

\section{Semantic Port-Aware Adaptive-Rate Transmission Protocol}

As illustrated in Fig.\ref{fig:Architecture}, the proposed Semantic Port-Aware Adaptive-Rate Transmission Protocol (SPAT) protocol establishes a port-aware SemCom architecture over wireless networks, which consists of three major components: transmitter, wireless channel, the receiver. The SPAT framework enables port-aware SemCom by jointly embedding transport-layer port information into semantic representations and performing adaptive semantic transmission over wireless channels.

\subsection{System Model}

At the transmitter, the source data $\bm{s}$ is processed by the semantic encoder to extract the compact semantic representations. Formally, the semantic encoding process can be expressed as
\begin{equation}
    \bm{x} = f_{\text{enc}}(\bm{s} \, ; \, \bm{\theta_{\textbf{enc}}}),
    \label{eq:enc}
\end{equation}
where $f_{\text{enc}}(\, \bm{\cdot} \,;\, \bm{\theta_{\textbf{enc}}} \,)$ denotes the semantic encoding function with the trainable parameters $\bm{\theta_{\textbf{enc}}}$, and $\bm{x}$ represents the extracted semantic feature.

To enhance the robustness of port identification while reducing header overhead, the proposed SPAT protocol adopts a joint port-payload embedding mechanism. Specifically, the source port $p_s$ and destination port $p_d$ are embedded into high-dimensional representations and then fused with the semantic feature $\bm{x}$ through the joint port-payload encoder. The process can be express as
\begin{equation}
    \bm{u} = f_{\text{joint}}(\bm{x}, \, p_s, \, p_d; \, \bm{\psi_{\textbf{joint}}}),
    \label{eq:portencoder}
\end{equation}
where $f_{\text{joint}}(\, \cdot \, ; \, \bm{\psi_{\textbf{joint}}})$ denotes the joint port-payload encoding function with the trainable parameters $\bm{\psi_{\textbf{joint}}}$, and $\bm{u}$ denotes the port-aware semantic representation.

To further improve transmission efficiency under varying channel conditions, an adaptive-rate controller is incorporated. Based on the channel state information (CSI) fed back from the receiver, the controller dynamically adjusts the semantic coding rate according to the semantic importance ranking. The detailed network architecture is presented in Section V.

Subsequently, the encoded semantic features are quantized, interleaved, packetized, and digitally modulated to generate the transmitted signal $\bm{y}$, which is then delivered over the wireless channel. The received signal $\bm{\hat{y}}$ can be expressed as
\begin{equation}
    \bm{\hat{y}} = \sqrt{P} \bm{H} \bm{y} + \bm{n},
    \label{eq:channelpass}
\end{equation}
where $\bm{H}$ denotes the channel matrix, $\bm{n}$ is the additive white Gaussian noise (AWGN) with $\bm{n} \sim \mathcal{CN}(0, \sigma_n^2 \bm{I})$, and $P$ is the thransmit power. The receiver is assumed to have perfect CSI for signal equalization. Based on the equalized signal, the post-equalization SNR is estimated, which is fed back to the transmitter for adaptive rate control.

\textbf{Note:} Although \eqref{eq:channelpass} is commonly adopted in SemCom, the proposed SPAT framework further takes into account the impact of transport-layer port information on the transmission process. Specifically, errors in the port-related representations may degrade the accuracy of port identification at the receiver, thereby potentially causing packet loss.

At the receiver, the SPAT framework supports two transmission scenarios, namely uplink and downlink semantic transmission. The received signal is first digitally demodulated, de-packetized, de-interleaved, and de-quantized. The recovered semantic representation is then fed into a port identification module to estimate the corresponding port information.

\textbf{Note:} For the uplink, both the source port and destination port are identified to determine the associated service. For the downlink, only the source port is required for port identification, while the destination port is utilized to enable conditional decoding, ensuring that semantic data are decoded only by the intended destination terminal.

In the uplink, the base station (BS) directly decodes the recovered semantic representation to reconstruct the source content:
\begin{equation}
    \bm{\hat{s}} = f_{\text{dec}}(\bm{\hat{x}}; \, \bm{\theta_{\textbf{dec}}}),
    \label{eq:dec}
\end{equation}
where $f_{\text{dec}}(\, \cdot \, ; \, \bm{\theta_{\textbf{dec}}})$ denotes the semantic decoding function with the trainable parameters $\bm{\theta_{\textbf{dec}}}$. By contrast, in the downlink, the recovered semantic representation is first processed by a conditional gating module to select the decoder associated with the target user, which will be detailed in Section IV-C. The gated features are subsequently decoded to reconstruct the desired content.

\begin{figure*}[t]
\centering
\includegraphics[width=0.98\textwidth]{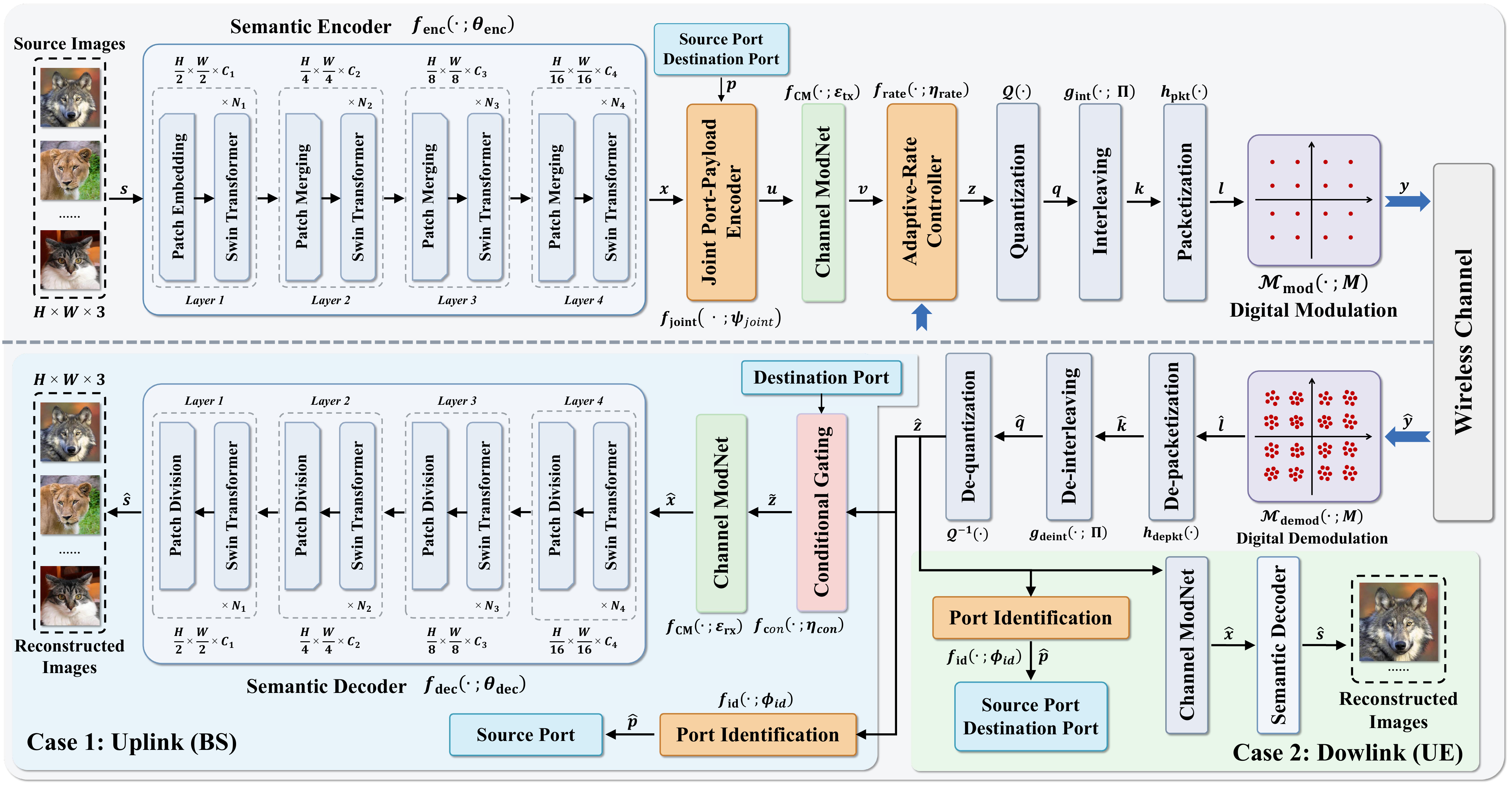}
\caption{The overall architecture of the proposed STAP-based digital semantic communication framework, illustrating the asymmetric processing mechanisms in uplink and downlink transmissions.}
\label{fig:UplinkandDownlink}
\end{figure*}

\subsection{Optimization Objective}

To train the proposed SPAT framework, a unified loss function is adopted to jointly optimize semantic reconstruction, port identification, and rate adaptation. The overall objective is defined as
\begin{equation}
\mathcal{L} = \lambda_1 L_{\text{rec}} + \lambda_2 L_{\text{port}} + \lambda_3 L_{\text{rate}},
\label{eq:overallloss}
\end{equation}
where $L_{\text{rec}}$, $L_{\text{port}}$, and $L_{\text{rate}}$ denote the semantic reconstruction loss, port identification loss, and adaptive-rate regularization term, respectively, and $\lambda_1$, $\lambda_2$, and $\lambda_3$ are predefined hyperparameters. By jointly minimizing the above objective, the SPAT framework learns to perform joint port-payload semantic encoding, port identification, adaptive-rate transmission, and conditional decoding, enabling efficient SemCom over wireless networks. The detailed implementation of the loss function is presented in the following section.

\section{Port-Aware Uplink And Downlink Semantic Transmission Framework}

In this section, based on the proposed SPAT protocol, we present a port-aware semantic transmission framework for both uplink and downlink scenarios. 

\subsection{The Overall Architecture}


The overall framework is illustrated in Fig.~\ref{fig:UplinkandDownlink}. In the semantic encoder, taking image datas as an example, the input image $\bm{s} \in \mathbb{R}^{H \times W \times 3}$ is first divided into non-overlapping patches and mapped into patch-level feature embeddings. These embeddings are then processed by Swin Transformer blocks to extract semantic information while preserving spatial structure. To obtain a more compact latent representation, patch merging is further applied to aggregate neighboring features through linear projection, thereby reducing the token resolution \cite{yang2024swinjscc}. 

\begin{figure*}[t]
\centering
\includegraphics[width=0.93\textwidth]{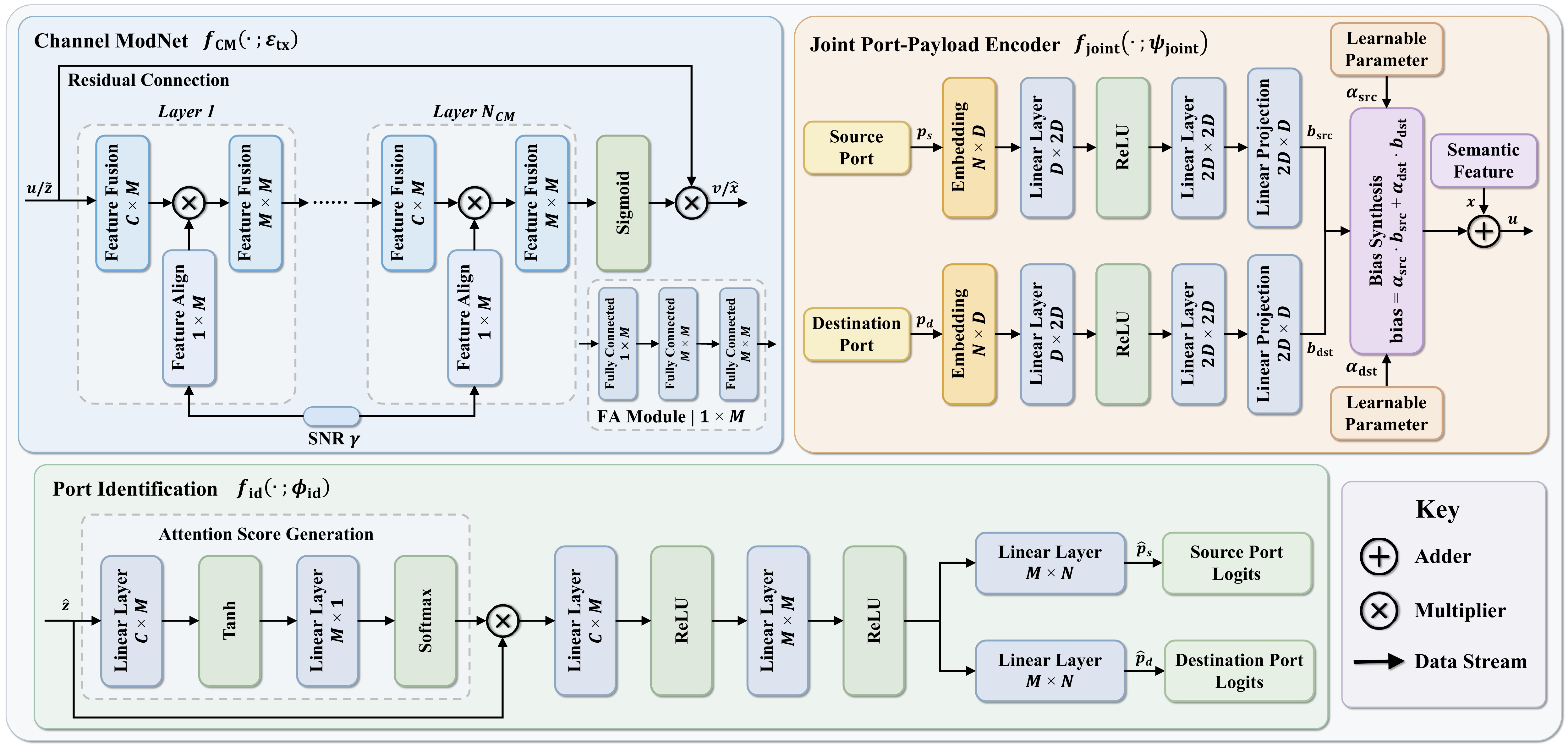}
\caption{The network architectures of the proposed Channel ModNet, Joint Port-Payload Encoder, and Port Identification modules.}
\label{fig:modules}
\end{figure*}

Subsequently, the semantic representation $\bm{x} \in \mathbb{R}^{M_t \times 1}$ is fused with the the source and destination port information $\bm{p}= \{p_s, \, p_d\}$ through the joint port-payload encoder to generate the port-aware semantic feature $\bm{u} \in \mathbb{R}^{M_t \times 1}$. The process can be expressed in \eqref{eq:portencoder}. The feature $\bm{u}$ is further processed by the CSI-aware Channel ModNet to produce the refined semantic feature $\bm{v}$. Specifically, as shown in Fig.~\ref{fig:modules}, the Channel ModNet consists of a Feature Alignment (FA) module and a Feature Fusion (FU) module, which jointly incorporate the signal-to-noise ratio (SNR) into the semantic representation., which can be expressed as
\begin{equation}
    \bm{v} = f_{\text{CM}}(\bm{u}, \, \gamma \,;\, \bm{\varepsilon_{\textbf{tx}}}), \,\,\,\, \bm{v} \in \mathbb{R}^{M_t \times 1},
    \label{eq:CMimgs}
\end{equation}
where $f_{\text{CM}}(\, \bm{\cdot} \,;\, \bm{\varepsilon_{\textbf{tx}}} \,)$ represents the Channel ModNet function with the trainable parameters $\bm{\varepsilon_{\textbf{tx}}}$, and $\gamma$ represents SNR. 

Subsequently, to further improve transmission efficiency, the adaptive-rate controller takes the semantic feature $\bm{v}$ and the CSI as inputs, and dynamically selects the most important features to generate the compressed representation $\bm{z}$
\begin{equation}
    \bm{z} = f_{\text{rate}}(\bm{v}, \, \gamma; \, \bm{\eta_{\textbf{rate}}}), \,\,\,\, \bm{z} \in \mathbb{R}^{\hat{M}_t \times 1},
    \label{eq:ratecontroller}
\end{equation}
where $\hat{M}_t \leq M_t$, $f_{\text{rate}}(\, \cdot \, ; \, \bm{\eta_{\textbf{rate}}})$ denotes the adaptive-rate mapping function with the trainable parameters $\bm{\eta_{\textbf{rate}}}$. 

Inspired by \cite{gong2025digital}, a quantization process is applied to discretize continuous values into integer symbols without introducing any learnable parameters. Specifically, the feature values are first normalized to the range $[0,1]$, and then a rounding operation is performed to produce discrete integer symbols. Subsequently, each integer symbol is converted into bit representation. The process can be formulated as
\begin{equation}
    \bm{q} = \mathcal{Q}(\bm{z}), \,\,\,\, \bm{q} \in \{0,1\}^{\hat{M}_t \cdot 2^{b} \times 1}
    \label{eq:quan}
\end{equation}
where $b$ denotes the number of bits per symbol, and $\mathcal{Q}(\cdot)$ denotes the quantization function.

The binary signal $\bm{q}$ is subsequently processed by a semantic-level interleaver, which redistributes the feature elements. The interleaving process can be expressed as:
\begin{equation}
    \bm{k} = g_{\text{int}}(\bm{q_1}, \cdots , \bm{q_{N}} \; ; \; \bm{\Pi}), \,\,\,\, \bm{k} \in \{0,1\}^{\hat{M}_t \cdot 2^{b} \times 1}.
    \label{eq:intlvimgs}
\end{equation}
where $g_{\text{int}}(\, \bm{\cdot} \, ; \, \bm{\Pi})$ denotes the interleaving function, $\bm{\Pi}$ denotes the semantic interleaving index vector, representing the permutation order of elements, and $\bm{k}$ denotes the interleaved feature.

\textbf{Note:} At the transport layer, the proposed SPAT framework adopts a port-embedding mechanism that considers only the header information related to port numbers. Other header fields, such as the segment length, can be incorporated in a similar manner but are omitted here for simplicity.

After interleaving, the semantic features $\bm{k}$ are packetized at the transport layer, forming a sequence of transmitted packets denoted by $\bm{l} \in \{0, \, 1\}^{T \times L}$, where $T \triangleq \lceil \frac{\hat{M}_t \cdot 2^{b} \times 1}{L} \rceil$ denotes the total number of packets and $L$ is the packet length. Formally, the packetization process can be expressed as
\begin{equation}
\bm{l} = h_{\text{pkt}}(\bm{k}), \,\,\,\, \bm{l} \in \{0, \, 1\}^{T \times L},
\label{eq:packetization}
\end{equation}
$h_{\text{pkt}}(\cdot)$ denotes the packetization function.

After packetization, the segmented data $\bm{l}$ are modulated to generate the constellation symbols $\bm{y}$. The digital modulation process can be expressed as
\begin{equation}
\bm{y} = \mathcal{M}_{\text{mod}}(\bm{l} \, ; \, M),
\label{eq:mod}
\end{equation}
where $\mathcal{M}(\cdot  \, ; \, M)$ represents the modulation function, $M$ denotes the modulation order (e.g., $M$-QAM), and $\bm{z} \in \mathbb{C}^{T \times L/log_{2}M}$ denotes the constellation symbols.The wireless channel characteristics are modeled as in \eqref{eq:channelpass}.

The received signal $\bm{\hat{y}}$ is first digitally demodulated and depacketized at the transport layer, and then processed through de-interleaving and dequantization to recover the semantic feature $\bm{\hat{z}}$. Formally, the process can be expressed as
\begin{equation}
    \bm{\hat{l}} = \mathcal{M}_{\text{demod}}(\bm{\hat{y}} \, ; \, M), \,\,\,\, \bm{\hat{l}} \in \{0, \, 1\}^{T \times L},
    \label{eq:demod}
\end{equation}
\begin{equation}
    \bm{\hat{k}} = h_{\text{depkt}}(\bm{\hat{l}}), \,\,\,\, \bm{\hat{k}} \in \{0,1\}^{\hat{M}_t \cdot 2^{b} \times 1},
    \label{eq:unpkt}
\end{equation}
\begin{equation}
    \bm{\hat{z}} = \mathcal{Q}^{-1}(g_{\text{deint}}(\bm{\hat{k}} \, ; \, \bm{\Pi})), \,\,\,\, \bm{\hat{z}} \in \mathbb{R}^{\hat{M}_t \times 1},
    \label{eq:unintlv}
\end{equation}
where $\mathcal{M}_{\text{demod}}(\cdot)$ denotes the digital demodulation function, $g_{\text{depack}}(\cdot)$ denotes the depacketization function, $g_{\text{deint}}(\cdot,;\bm{\Pi})$ denotes the de-interleaving operation that restores the original order according to the interleaving index matrix $\bm{\Pi}$, and $\mathcal{Q}^{-1}(\cdot)$ denotes the de-quantization function.

The recovered features are first restored to the original dimension $\mathbb{R}^{\hat{M}_t \times 1}$. Subsequently, the receiver operates under two communication scenarios, namely the uplink to the BS and the downlink to the user terminal (UE). For the downlink scenario, the receiver first identifies the source of the transmitted data. To this end, the recovered feature $\bm{\hat{z}}$ is fed into the port identification module to estimate the source port, which can be expressed as
\begin{equation}
    \hat{p}_s = f_{\text{id}}(\bm{\hat{z}} \, ; \, \bm{\phi_{\text{id}}}),
    \label{eq:upident}
\end{equation}
where $f_{\text{id}}(\, \cdot \, ; \, \bm{\phi_{\textbf{id}}})$ denotes the port identification function with the trainable parameters $\bm{\phi_{\textbf{id}}}$. In parallel, the recovered feature $\bm{\hat{z}}$ is further processed by a conditional gating module under the guidance of the destination port $p_d$, thereby selectively activating the semantic decoder corresponding to the intended UE. The gating operation is given by
\begin{equation}
    \bm{\tilde{z}} = f_{\text{con}}(\bm{\hat{z}}, \, p_d; \, \bm{\eta_{\textbf{con}}}), \,\,\,\, \bm{\tilde{z}} \in \mathbb{R}^{M_t \times 1},
    \label{eq:condition}
\end{equation}
where $f_{\text{con}}(\, \cdot \, ; \, \bm{\eta_{\textbf{con}}})$ denotes the port-aware gating function with the trainable parameters $\bm{\eta_{\textbf{con}}}$. The gated semantic features $\bm{\tilde{z}}$ is subsequently refined by the receiver-side Channel ModNet, which can be expressed as
\begin{equation}
    \bm{\hat{x}} = f_{\text{CM}}(\bm{\tilde{z}}, \, \gamma \,;\, \bm{\varepsilon_{\textbf{rx}}}), \,\,\,\, \bm{\hat{x}} \in \mathbb{R}^{M_t \times 1},
    \label{eq:CMimgsrx}
\end{equation}
where $f_{\text{CM}}(\cdot \,;\, \bm{\varepsilon_{\textbf{rx}}})$ represents the Channel ModNet mapping function with the trainable parameters $\bm{\varepsilon_{\textbf{rx}}}$. The refined semantic feature $\bm{\hat{x}}$ is then decoded by the semantic decoder to reconstruct the target content, as defined in \eqref{eq:dec}.

\begin{figure*}[t]
    \centering
    \includegraphics[width=0.9\textwidth]{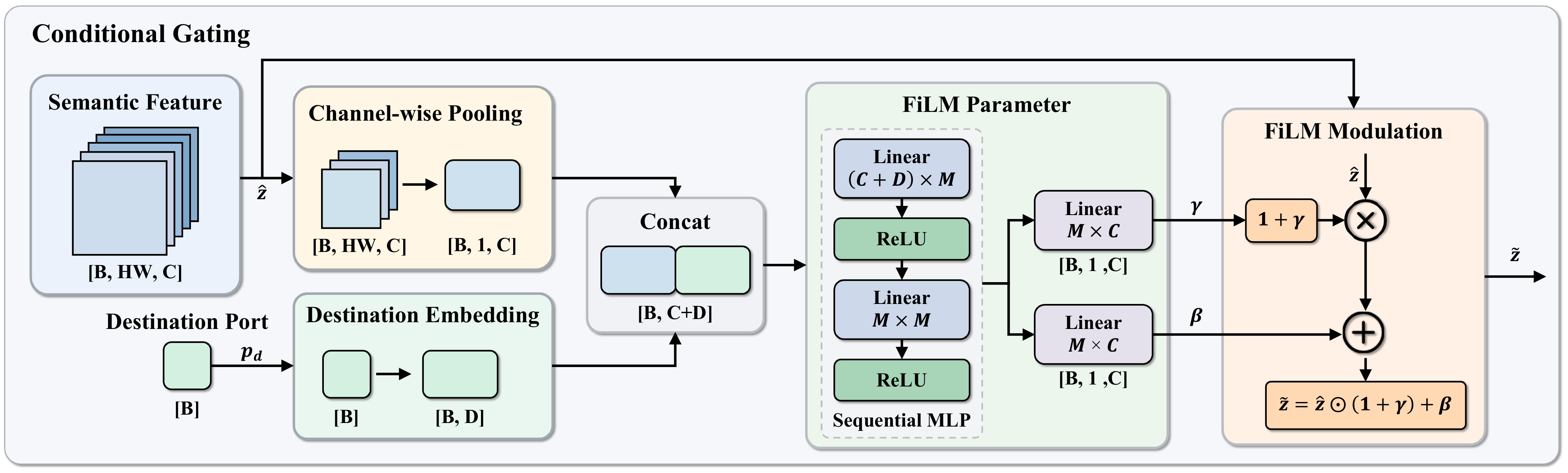}
    \caption{The network architecture of the proposed Conditional Gating module for destination-aware selective semantic decoding.}
    \label{fig:gating}
\end{figure*}

The network parameters are optimized by minimizing the overall objective in \eqref{eq:overallloss}. For the downlink scenario, the reconstruction loss is designed to ensure accurate recovery at the intended destination while suppressing undesired decoding at non-target terminals, which is defined as
\begin{equation}
    L_{\text{rec-down}} = \mathbb{E} \left[ d(\bm{s}, \, \bm{\hat{s}^{+}}) \right] + \mathbb{E} \left[ d(\bm{s_{\text{null}}}, \, \bm{\hat{s}^{-}}) \right],
    \label{eq:downrecloss}
\end{equation}
where $\bm{\hat{s}^{+}}$ and $\bm{\hat{s}^{-}}$ denote the reconstructed outputs at the intended and non-target terminals, respectively, $\bm{s_{\text{null}}}$ represents a null reference image, and $d(\cdot)$ measures the discrepancy between the target and reconstructed data, which is implemented as the mean squared error (MSE) in this work. Furthermore, the port identification loss in the downlink considers only the source port, which is defined as
\begin{equation}
    L_{\text{port-down}} = \mathbb{E} \left[ \ell(p_s, \, \hat{p}_s) \right],
    \label{eq:downportloss}
\end{equation}
where $\ell(\cdot)$ denotes the cross-entropy loss.

For the uplink scenario, the recovered feature $\bm{\hat{z}}$ is fed into the port identification module to estimate both the source port and the destination port, which is different from the downlink case. This process can be expressed as
\begin{equation}
    \{ p_s, \, p_d \} = f_{\text{id}}(\bm{\hat{z}}; \, \bm{\phi_{\textbf{id}}}).
    \label{eq:upportident}
\end{equation}
In parallel, the recovered feature $\bm{\hat{z}}$ is processed by the Channel ModNet and the semantic decoder to reconstruct the transmitted content, as given in \eqref{eq:CMimgsrx} and \eqref{eq:dec}. It is worth noting that incorrect port estimation may cause service misidentification, which is treated as an effective packet loss event.

Unlike the downlink case, the reconstruction loss is only required to ensure accurate recovery of semantic information, while the port identification loss jointly accounts for both the source port and the destination port, which are defined as
\begin{align}
     L_{\text{rec-up}} & = \mathbb{E} \left[ d(\bm{s}, \, \bm{\hat{s}}) \right], \label{uploss1} \\
    L_{\text{port-up}} & = \mathbb{E} \left[ \ell(p_s, \, \hat{p}_s) \right] + \mathbb{E} \left[ \ell(p_d, \, \hat{p}_d) \right]. \label{uploss2}
\end{align}

\subsection{Port Embedding and Identification}

To enable transport-aware semantic transmission, the proposed SPAT framework introduces a joint port-payload encoding and identification mechanism, as shown in Fig.~\ref{fig:modules}. Specifically, the transport-layer source and destination port information is embedded into the semantic feature at the transmitter, while the corresponding port attributes are inferred from the recovered semantic representation at the receiver, thereby enabling robustness to bit errors.

At the transmitter, each port index is first mapped into a learnable embedding space. The corresponding embedding vectors $\{\bm{e_s}, \, \bm{e_d}\}$ are given by
\begin{equation}
     \{\bm{e_s}, \, \bm{e_d}\} = f_{\text{PE}}(p_s, \, p_d; \, \bm{\psi_{\textbf{PE}}}), \,\,\,\, \{\bm{e_s}, \, \bm{e_d}\} \in \mathbb{R}^{D \times 1}, 
     \label{portembedding1}
\end{equation}
where $f_{\text{PE}}(\, \cdot \, ; \, \bm{\psi_{\textbf{PE}}})$ denotes the embedding function with the parameters $\bm{\psi_{\textbf{PE}}}$, and $D$ denotes the embedding dimension.

Subsequently, the port embeddings $\{\bm{e_s}, \, \bm{e_d}\}$ are separately processed by lightweight multilayer perceptrons (MLPs). As shown in Fig.~\ref{fig:modules}, each branch consists of two linear layers with a ReLU activation, followed by a linear projection. The resulting port-specific representations $\{\bm{b_{\textbf{src}}}, \, \bm{b_{\textbf{dst}}}\}$ are then fused into a unified port-aware bias term. Given the semantic feature $\bm{x}$, the joint port-payload encoder incorporates the synthesized port bias through an adder, yielding the port-aware semantic feature $\bm{u}$. The process is expressed as
\begin{align}
     \textbf{bias} & = \bm{\alpha_{\text{src}}} \cdot \bm{b}_{\text{src}} + \bm{\alpha_{\text{dst}}} \cdot \bm{b}_{\text{dst}}, 
     \label{eq:bias_syn} \\
    \bm{u} & = \bm{x} + \textbf{bias},
    \label{eq:joint_add}
\end{align}
where $\textbf{bias}$ denotes the port-aware bias injected into the semantic payload feature.

\textbf{Note:} Compared with directly appending explicit header fields, the proposed port-embedding mechanism provides two main advantages. It reduces protocol overhead by avoiding bit-level transmission of port metadata, and improves robustness to channel distortion by enabling the receiver to infer port attributes from the recovered semantic representation.

At the receiver, a port identification module is introduced to infer the embedded port information $\{ \hat{p}_s, \, \hat{p}_d\}$ from the recovered semantic feature $\bm{\hat{z}}$, as shown in Fig.~\ref{fig:modules}. Specifically, the recovered feature $\tilde{\bm{z}}$ is first fed into an attention-score generation block to emphasize port-sensitive dimensions, and the resulting attention is applied to the feature through element-wise multiplication. The attended feature is then processed by MLP to obtain a compact latent representation for port inference. Finally, two parallel classifiers are employed to estimate the source-port and destination-port logits:
\begin{equation}
    \{\bm{\hat{o}_s} \, \bm{\hat{o}_d}\} = f_{\text{id}}(\bm{\hat{z}} \, ; \, \bm{\phi_{\text{id}}}), \,\,\,\, \{\bm{\hat{o}_s}, \, \bm{\hat{o}_d}\} \in \mathbb{R}^{N \times 1},
    \label{eq:port_logits1}
\end{equation}
where $\bm{\hat{o}_s}$, $\bm{\hat{o}_d}$ denote the source-port and destination-port logits, respectively. The predicted ports are obtained as
\begin{equation}
    \hat{p}_s = \arg\max \, \bm{\hat{o}_s}, \qquad
    \hat{p}_d = \arg\max \, \bm{\hat{o}_d}.
    \label{eq:port_pred}
\end{equation}
The proposed conditional gating mechanism further suppresses unintended semantic reconstruction across non-target ports.

\subsection{Conditional Gating Mechanism}

To enable destination-aware selective semantic decoding in the downlink, the proposed SPAT framework introduces a conditional gating mechanism. This module modulates the recovered semantic feature according to the destination-port condition, so that only the intended UE can activate the effective semantic reconstruction path. 

First, as shown in Fig.~\ref{fig:gating}, the recovered semantic feature $\hat{\bm{z}}$ is processed by a channel-wise pooling operator to obtain a compact global descriptor that aggregates spatial information while preserving channel-wise semantic statistics. Meanwhile, the destination port is mapped into a learnable embedding, which is then concatenated with the pooled semantic descriptor to form a joint condition vector. The concatenated condition vector is then fed into a lightweight MLP to generate the Feature-wise Linear Modulation (FiLM) parameters, namely the scaling factor $\bm{\gamma}$ and the shifting factor $\bm{\beta}$.

Finally, the recovered semantic feature $\hat{\bm{z}}$ is modulated in a feature-wise manner according to the generated FiLM parameters $\{\bm{\gamma}, \, \bm{\beta}\}$. The modulation process is given by
\begin{equation}
    \tilde{\bm{z}} = \hat{\bm{z}} \odot (1 + \bm{\gamma}) + \bm{\beta}, \,\,\,\, \bm{\tilde{z}} \in \mathbb{R}^{M_t \times 1},
    \label{eq:film_modulation}
\end{equation}
where $\odot$ denotes element-wise multiplication.

The conditional gating mechanism enables the destination-port condition to directly modulate the semantic representation before decoding. When the receiver matches the intended destination, informative semantic components are preserved for accurate reconstruction; otherwise, irrelevant components are suppressed to reduce unintended decoding. In this way, the proposed module establishes a destination-aware pathway for selective semantic decoding in the downlink.

\section{Adaptive-Rate Control Mechanism}

To improve transmission efficiency under varying channel conditions, the proposed SPAT framework introduces an adaptive-rate control mechanism, as shown in Fig.~\ref{fig:Adaptive_Rate_Controller}. This module dynamically adjusts the number of transmitted semantic channels according to channel conditions and feature importance, thereby achieving a better tradeoff between transmission reliability and spectral efficiency. Given the port-aware semantic feature $\bm{v}$, the adaptive-rate controller generates a compressed semantic representation $\bm{z}$ by selecting the most important semantic channels, as expressed in \eqref{eq:ratecontroller}.

\subsection{SNR-Based Rate Estimation}

To adapt the transmission rate to channel conditions, the number of retained semantic channels is first determined according to the input SNR. Let $C_r$ denote the number of selected channels, where $C_{\min} \leq C_r \leq C_{\max}$, $C_{\text{max}}$ is the channel dimension of the semantic feature $\bm{v}$, and $C_{\text{min}}$ is a predefined minimum channel budget. Inspired by a smooth logistic mapping, $C_r$ is determined as
\begin{equation}
    C_r = \left \lfloor C_{\max} - s(\gamma) (C_{\max} - C_{\min}) \right \rceil,
    \label{eq:Cr_def}
\end{equation}
where
\begin{equation}
    s(\gamma) = \frac{1}{1 + \exp ( - \frac{\gamma - \gamma_{\text{mid}}} {\tau})},
    \label{eq:snr_sigmoid}
\end{equation}
and
\begin{equation}
    \gamma_{\text{mid}} = \frac{\gamma_{\text{low}} + \gamma_{\text{high}}}{2},
    \label{eq:snr_mid}
\end{equation}
where $\left \lfloor \right \rceil$ represents the rounding operation, $\gamma_{\text{low}}$ and $\gamma_{\text{high}}$ denote the lower and upper SNR thresholds, respectively, and $\tau$ is a smoothing factor controlling the transition sharpness. According to \eqref{eq:Cr_def}-\eqref{eq:snr_mid}, a lower SNR yields a larger $C_r$, so that more semantic channels are retained to enhance robustness; conversely, a higher SNR leads to a smaller $C_r$, thereby improving transmission efficiency.

\begin{figure}[t]
    \centering
    \includegraphics[width=0.48\textwidth]{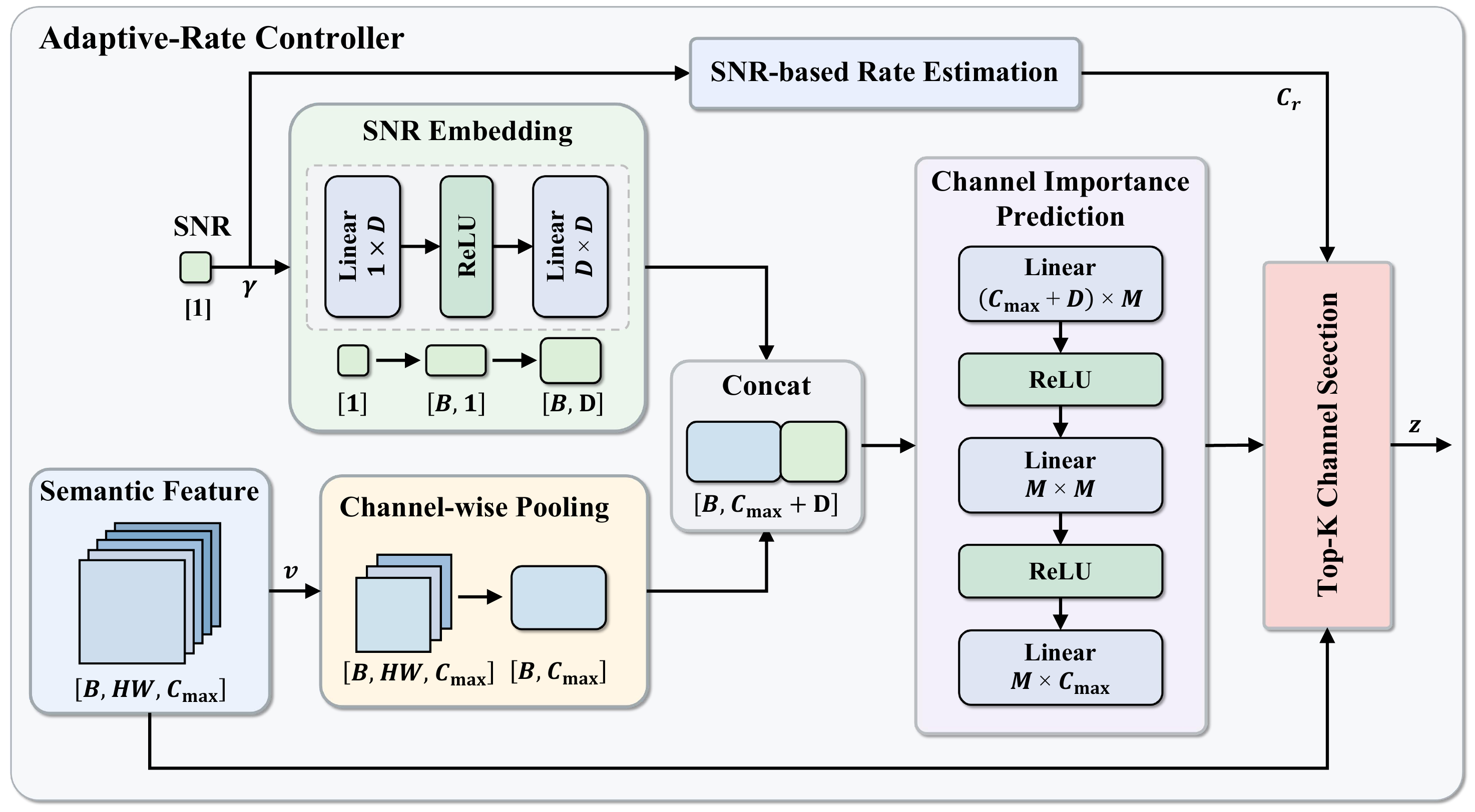}
    \caption{The network architecture of the Adaptive-Rate Controller.}
    \label{fig:Adaptive_Rate_Controller}
\end{figure}

\subsection{Channel Importance Prediction}

The proposed mechanism further selects the most informative semantic channels based on feature importance. Specifically, the input semantic feature $\bm{v}$ is first compressed by channel-wise pooling, while the input SNR $\gamma$ is mapped to a learnable embedding. The resulting representations are concatenated and passed through a lightweight MLP to predict channel-importance scores, which are then transformed into soft channel weights $\bm{w}$ via sigmoid activation. Finally, according to the estimated channel budget $C_r$, a Top-$K$ operator retains the $C_r$ most important channels, yielding the compressed semantic feature $\bm{z}$ as
\begin{equation}
    \bm{z} = f_{\text{Top-K}}(\bm{v}, \bm{w}; C_r),
    \label{eq:topk_select}
\end{equation}
where \(\bm{w}\) denotes the predicted channel-importance logits. In this way, the adaptive-rate controller preserves the most informative semantic components according to the channel condition while reducing transmission redundancy.

\subsection{Adaptive-Rate Optimization}

To train the adaptive-rate controller in an end-to-end manner, an adaptive-rate regularization term is introduced. Specifically, the rate loss consists of a ratio constraint term and a binarization regularization term, which is defined as
\begin{equation}
    L_{\text{rate}} = \alpha L_{\text{ratio}} + \beta L_{\text{bin}},
    \label{eq:rateloss}
\end{equation}
where $\alpha$ and $\beta$ are predefined hyperparameter.

The ratio constraint $L_{\text{ratio}}$ encourages the average soft channel weight $\bm{w}$ to match the target transmission ratio $C_r$:
\begin{equation}
    L_{\text{ratio}} = \left( \frac{1} {C_{\max}} \sum_{i=1}^{C_{\max}} w_i - \frac{C_r} {C_{\max}} \right)^2,
    \label{eq:l_ratio}
\end{equation}
where $w_i$ denotes the importance weight of the $i$-th channel. In addition, the binarization regularization $L_{\text{bin}}$ promotes near-binary channel selection:
\begin{equation}
    L_{\text{bin}} = \frac{1}{C_{\max}} \sum_{i=1}^{C_{\max}} w_i(1-w_i),
    \label{eq:l_bin}
\end{equation}
which reaches its minimum when each $w_i$ approaches either $0$ or $1$, thereby yielding a clearer separation between selected and discarded channels. By jointly optimizing \eqref{eq:l_ratio} and \eqref{eq:l_bin}, the predicted channel-importance weights are encouraged to match the target channel budget while approaching near-binary selection, thereby enabling the adaptive-rate controller to retain the most informative semantic channels.

\textbf{Note:} The proposed adaptive-rate control mechanism provides two main advantages. It adjusts the number of transmitted semantic channels according to the channel SNR, and combines channel-importance prediction with Top-$K$ selection to preserve informative semantic components while reducing redundancy. As a result, the proposed mechanism enables SPAT to achieve adaptive semantic compression with improved robustness and transmission efficiency.

\section{Experimental Results and Analysis}

\subsection{Experimental Setup}

\textit{1) Datasets:} To validate the applicability of our SPAT framework for image reconstruction tasks, we utilized AFHQ dataset and ImageNet-10 dataset. The AFHQ dataset comprises 15,000 high-quality animal face images. The ImageNet-10 dataset consists of images selected from ten object categories of the ImageNet large-scale visual database. Both datasets are resized to $256 \times 256$ for reconstruction tasks.

\textit{2) Baseline Methods:} To comprehensively evaluate the performance of the proposed SPAT framework, several representative baseline methods are implemented for comparison:
\begin{itemize}
    \item {\textbf{TCP Scheme \cite{tian2005tcp}:} TCP is adopted as a benchmark for reliable transmission. The TCP-based framework is built upon the SwinJSCC architecture extended to the digital domain, which ensures bit-level integrity through ACK and retransmission mechanisms. A maximum retransmission limit is imposed instead of assuming infinite retries.}
    \item{\textbf{UDP Scheme \cite{gu2007udt}:} UDP is considered as the baseline for low-latency communication, which performs connectionless transmission without ACK or retransmissions. Similar to the TCP configuration, the UDP-based pipeline employs SwinJSCC with digital modulation.}
    \item{\textbf{SITP Scheme \cite{wang2025sitp}:} SITP is adopted as the SemCom baseline, which performs one-shot transmission without ACK or retransmissions. It verifies only port information while allowing corrupted semantic payloads to be delivered for reconstruction. The SITP-based pipeline employs SwinJSCC with digital modulation.}
\end{itemize}

\textit{3) Performance Metrics:} To assess the performance of the proposed model, we adopt both semantic-level and pixel-level mertics. For semantic evaluation, the Learned Perceptual Image Patch Similarity (LPIPS) metric is employed to calculate the perceptual similarity of the reconstructed images. LPIPS evaluates the perceptual difference between two images by comparing their deep feature representations extracted from a pretrained NN, with lower score indicates the higher perceptual similarity. For pixel-level evaluation, Peak Signal to Noise Ratio (PSNR) measures the fidelity of the reconstructed image, with higher values indicating reduced distortion. Multi-Scale Structural Similarity Index Measure (MS-SSIM) evaluates perceptual similarity by considering luminance, contrast, and structural information across multiple scales, where higher values signify greater similarity.

\begin{figure}[t]
    \centering
    \includegraphics[width=0.45\textwidth]{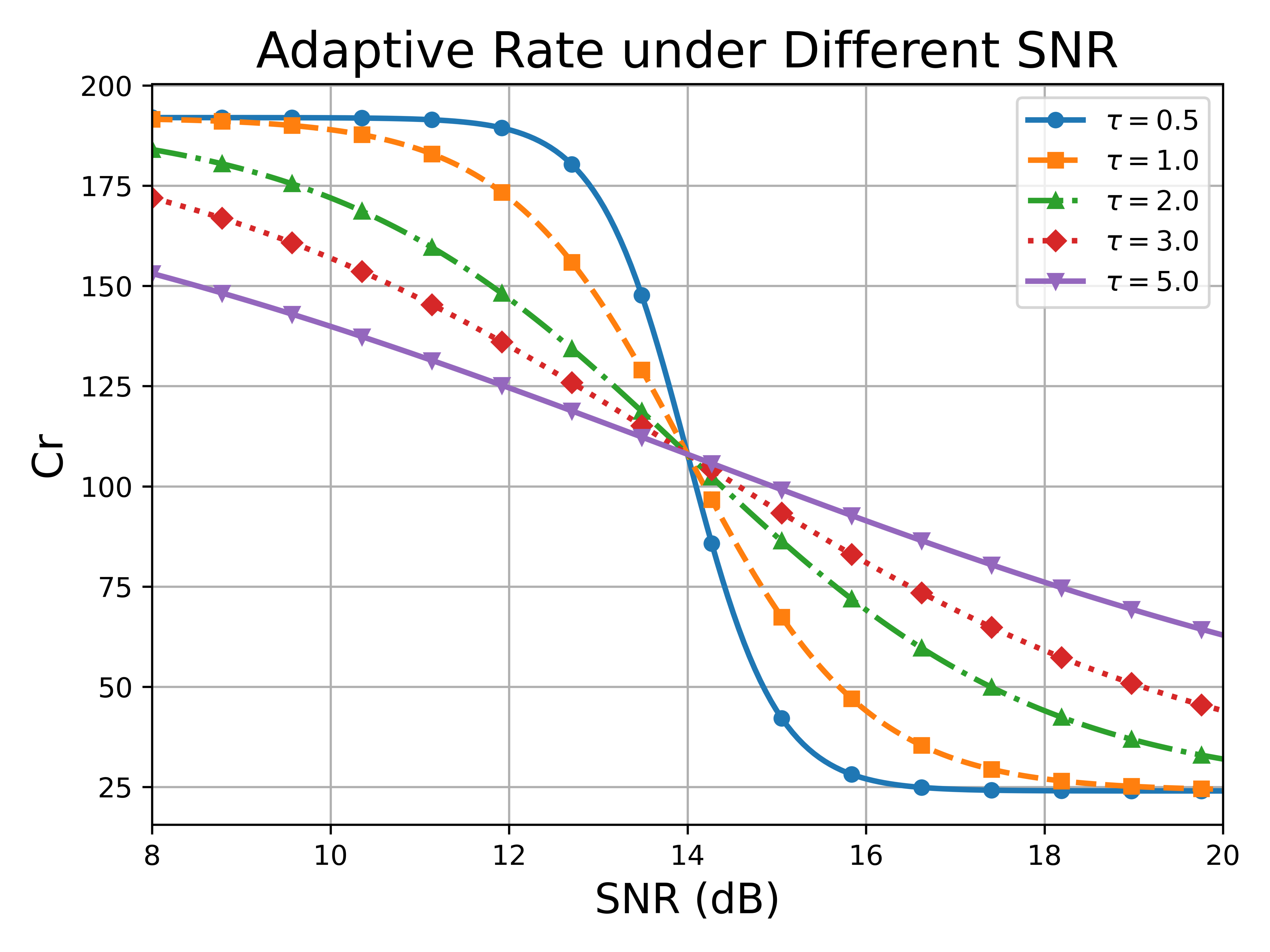}
    \caption{Adaptive-rate behavior of the proposed SPAT framework under different SNRs $\gamma$ and temperature coefficients $\tau$.}
    \label{fig:Adaptive_Rate_Controller}
\end{figure}

\captionsetup[table]{justification=centering, labelsep=space, textfont=sc} 
\begin{table}[t]
    \renewcommand\arraystretch{1.8}
    \setlength{\tabcolsep}{4pt}
    \caption{ \\ Parameter Setting of Adaptive-Rate Evaluation}
    \centering
    \begin{tabular}{cc}
        \hline
        \hline
        Parameters & Values \\
        \hline
        $C_{\text{min}}$ & 24  \\
        $C_{\text{max}}$ & 192 \\
        $\gamma_{\text{low}}$ & 8.0 \\
        $\gamma_{\text{high}}$ & 20.0 \\
        SNR Range (dB) & [8.0, 8.06, $\cdots$, 19.94, 20.0] \\
        Temperature Coefficient $\tau$ & [0.5, 1.0, 2.0, 3.0, 5.0] \\
        \hline
        \hline
    \end{tabular}
    \label{tab:adaptive_rate}
\end{table}

\subsection{Adaptive-Rate Evaluation}

To evaluate the adaptive-rate behavior of the proposed SPAT framework, we analyze the variation of the retained semantic channel number $C_r$ under different SNRs and temperature coefficients $\tau$. According to \eqref{eq:Cr_def}-\eqref{eq:snr_mid}, the adaptive-rate controller determines the channel budget by mapping the channel condition to the number of retained semantic channels. In this experiment, the SNR $\gamma$ varies from 8 dB to 20 dB, while multiple values of $\tau$ are considered to examine how the temperature coefficient affects the smoothness and sensitivity of rate adaptation. The detailed parameter settings are summarized in Table \ref{tab:adaptive_rate}.

Fig.~\ref{fig:Adaptive_Rate_Controller} shows the adaptive-rate curves under different values of $\tau$. For all cases, the retained channel number $C_r$ decreases monotonically as the SNR $\gamma$ increases, which is consistent with the design objective of the proposed mechanism: more semantic channels are preserved under poor channel conditions to improve robustness, whereas fewer channels are retained under favorable conditions to reduce redundancy and improve spectral efficiency. In addition, $\tau$ has a significant effect on the transition behavior of the adaptive-rate mapping. A smaller $\tau$ leads to a sharper transition, indicating more aggressive rate adaptation, while a larger $\tau$ produces a smoother and more gradual decrease in $C_r$, thereby avoiding abrupt rate changes.

\begin{figure*}[t]
    \centering
    \begin{minipage}{\textwidth}
        \centering
        \includegraphics[width=0.32\textwidth]{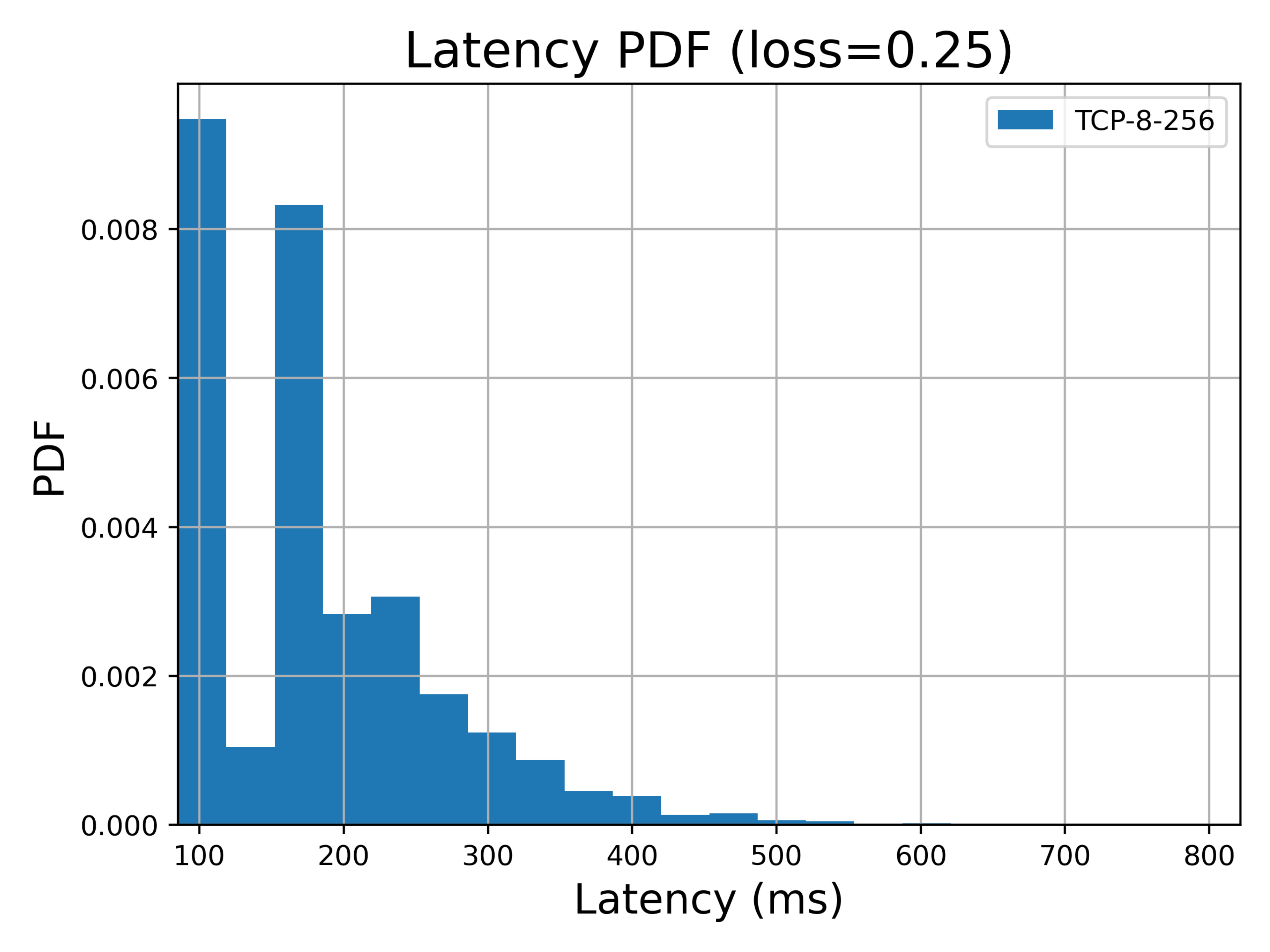} 
        \includegraphics[width=0.32\textwidth]{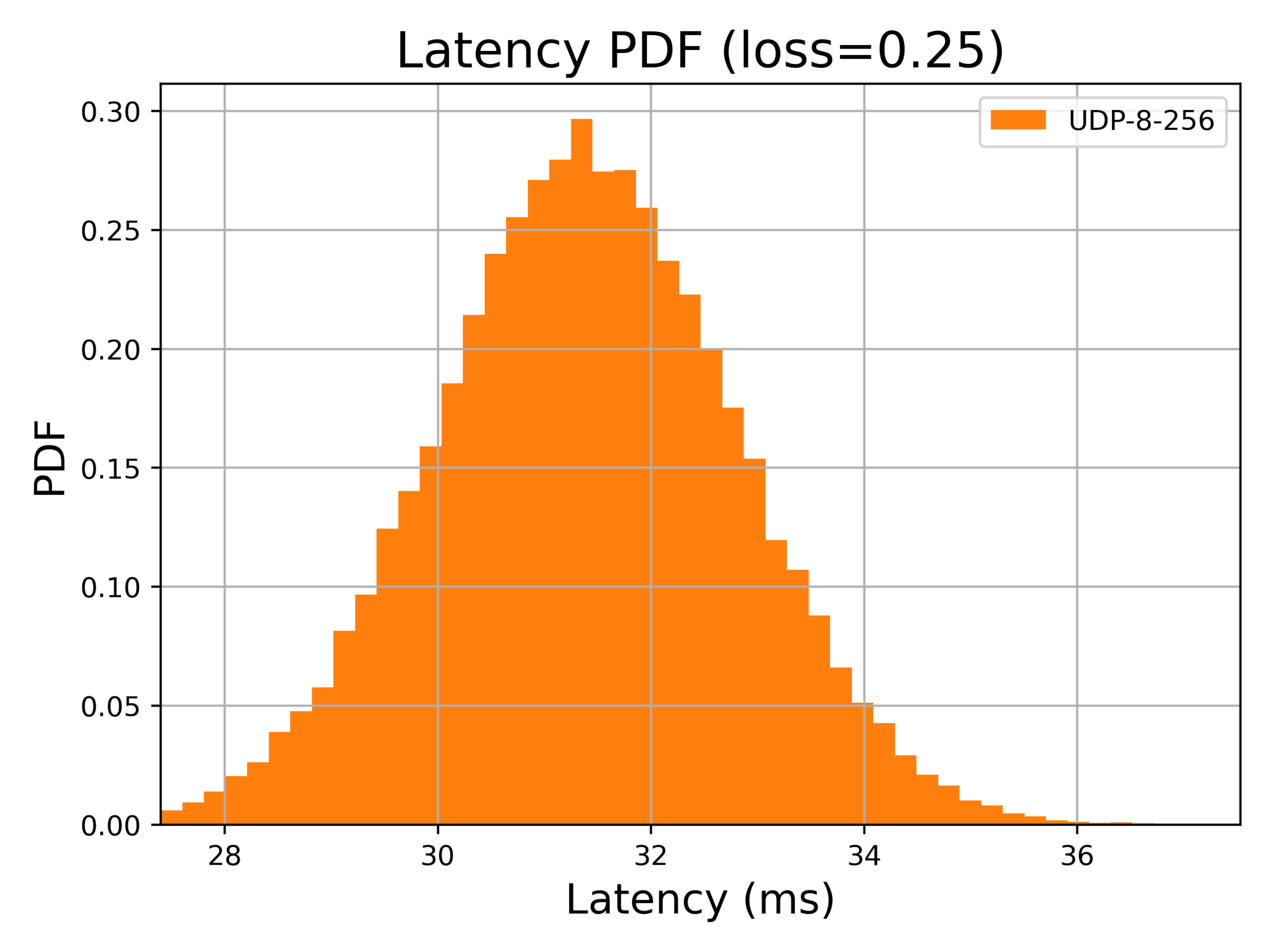} 
        \includegraphics[width=0.32\textwidth]{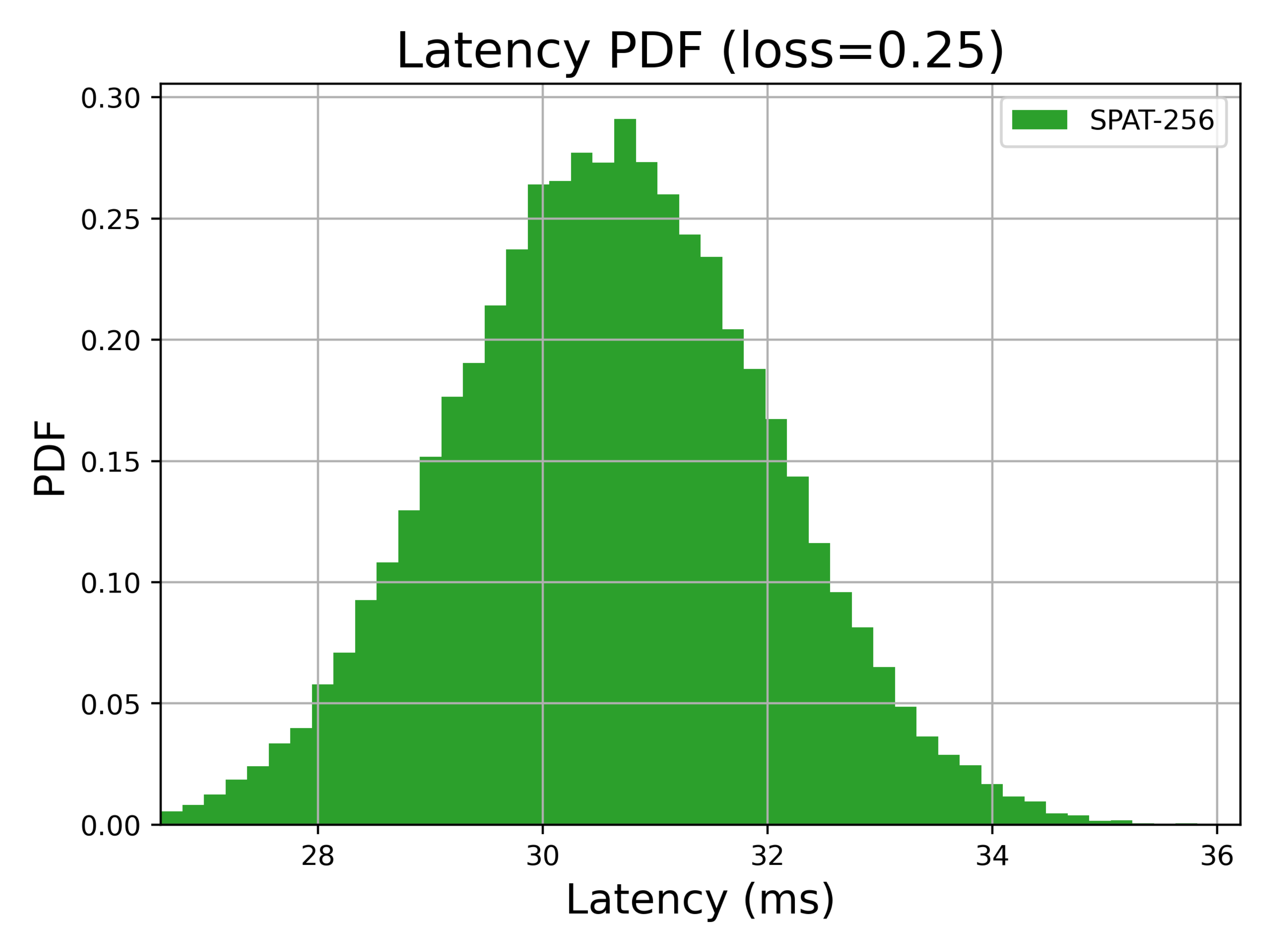} \\
        {\footnotesize (a) Latency PDFs.}
        \label{fig:latencya}
    \end{minipage}

    \vspace{2.0pt}

    \begin{minipage}{\textwidth}
        \centering
        \includegraphics[width=0.32\textwidth]{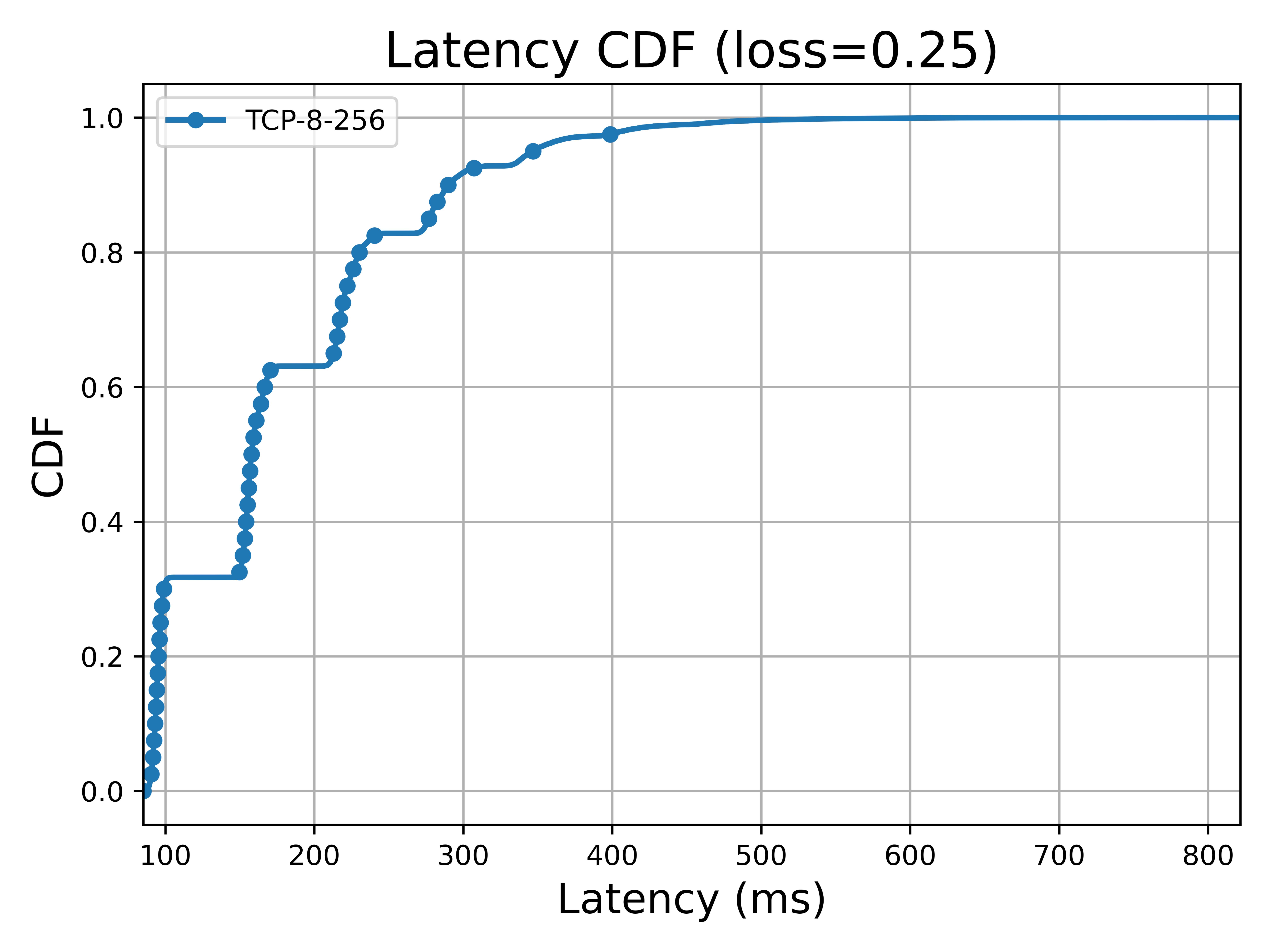}
        \includegraphics[width=0.32\textwidth]{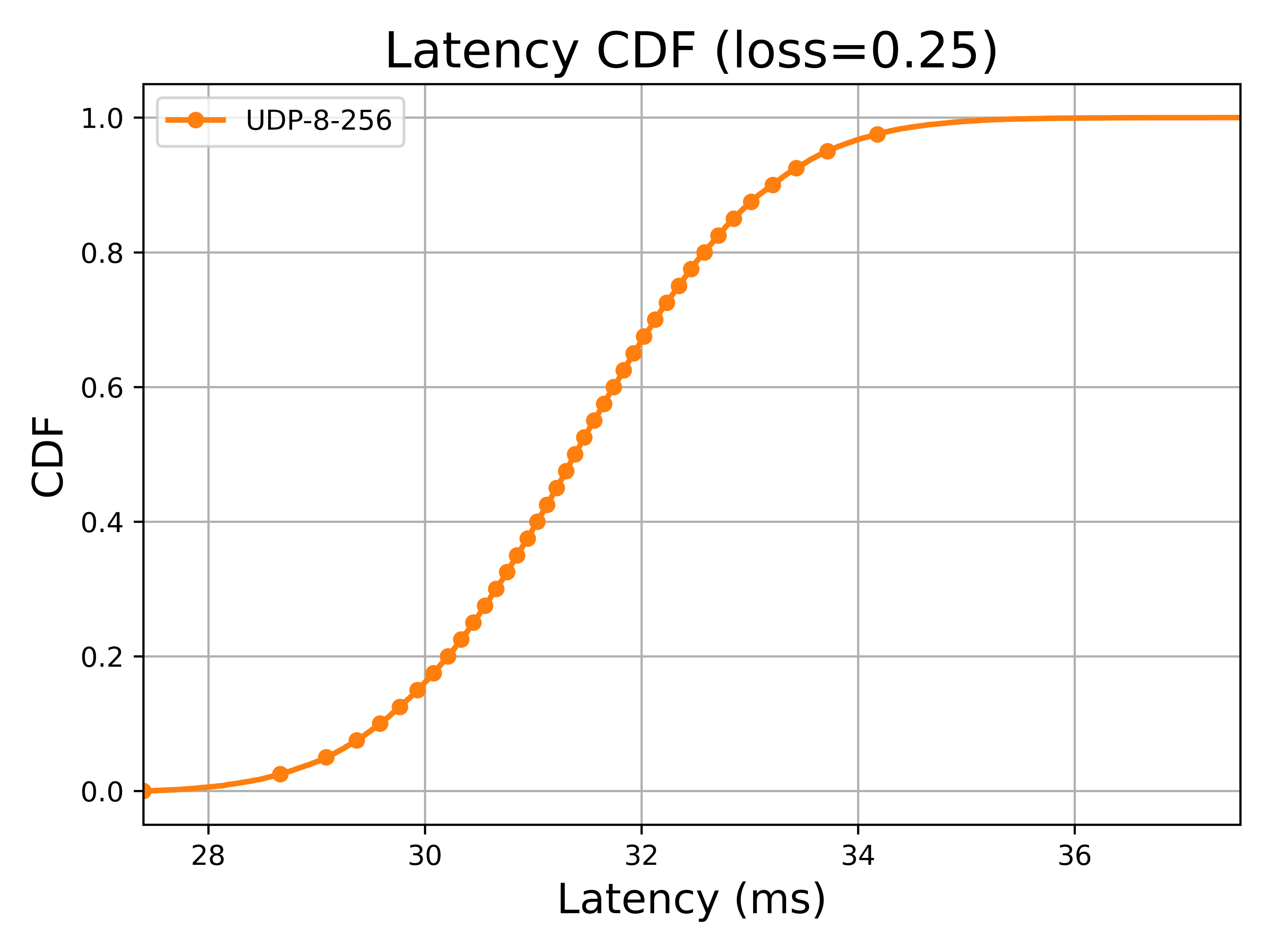}
        \includegraphics[width=0.32\textwidth]{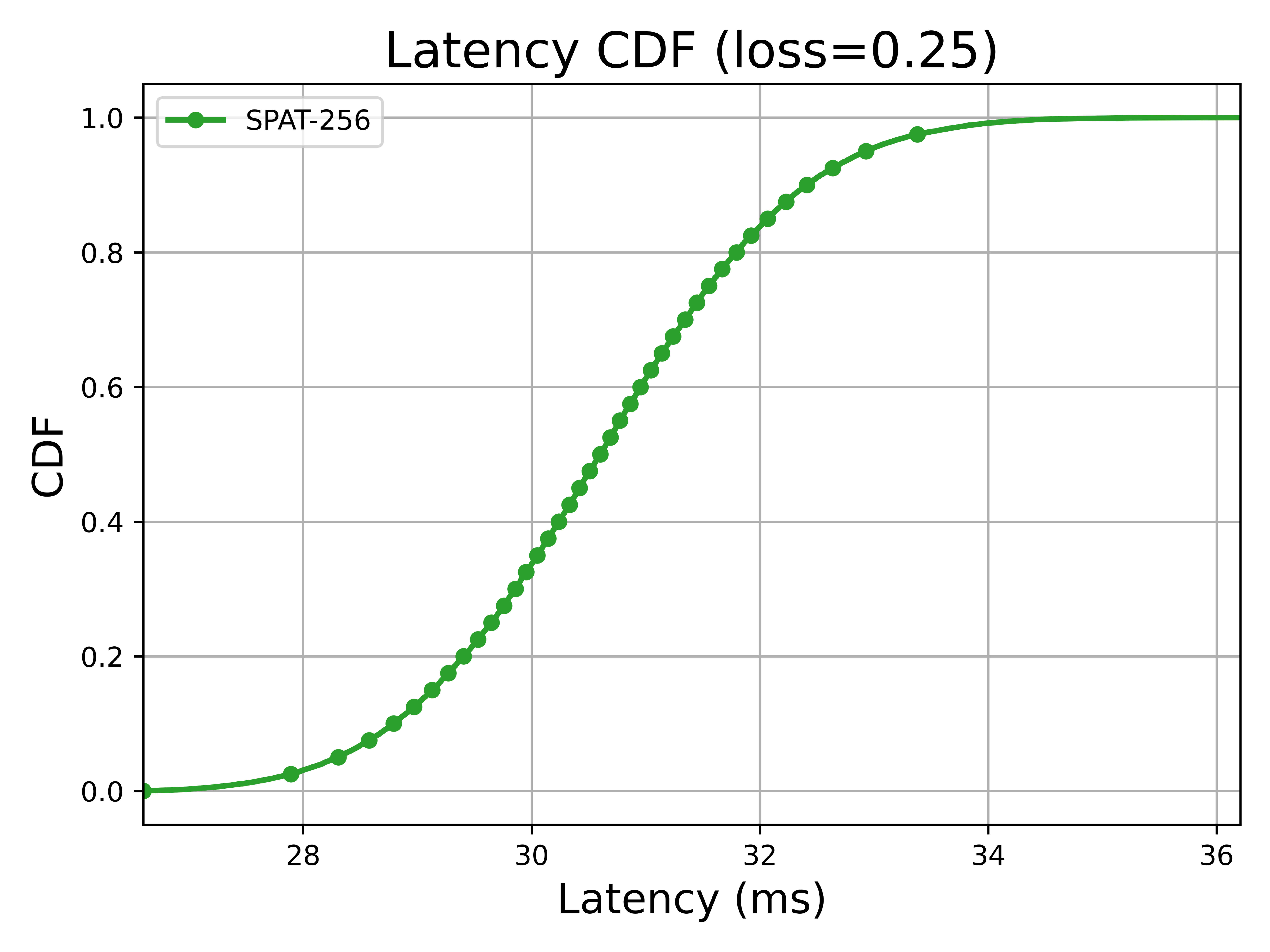} \\
        {\footnotesize (b) Latency CDFs.}
        \label{fig:latencyb}
    \end{minipage}

    \caption{Comparison of latency performance among TCP, UDP, and SPAT with a packet loss rate of 0.25. (a) illustrates the latency PDFs, while (b) presents the CDFs.}
    \label{fig:latency}
\end{figure*}

\captionsetup[table]{justification=centering, labelsep=space, textfont=sc} 
\begin{table}[t]
    \renewcommand\arraystretch{1.8}
    \setlength{\tabcolsep}{4pt}
    \caption{ \\ Parameter Setting of Latency Experiment}
    \centering
    \begin{tabular}{cc}
        \hline
        \hline
        Parameters & Values \\
        \hline
        Port / Payload Length & 8 / 256 bits \\
        Transmission Rate & 100 kbps \\
        Packet Loss Rate & 0.25 \\
        Retransmission Timeout & 40 ms \\
        Maximum TCP Retransmissions & 5 \\
        Mean One-way Delay & 5.0 ms\\
        One-way Delay Jitter & 1.414 ms \\
        Minimum One-way Delay & 1.0 ms \\
        \hline
        \hline
    \end{tabular}
    \label{latency}
\end{table}

\captionsetup[table]{justification=centering, labelsep=space, textfont=sc}
\begin{table}[t]
    \renewcommand\arraystretch{1.5}
    \setlength{\tabcolsep}{4.5pt}
    \caption{\\ Parameter Settings of Experiment}
    \centering
    \begin{tabular}{cccc}
        \hline
        \hline
        Parameters & Values & Parameters & Values \\
        \hline
        $\lambda_1$ & 1.0   & $\lambda_2$ & 10 \\
        $\lambda_3$ & 3.0   & $\tau$ & 3.0 \\
        $\alpha$    & 1.0   & $\beta$ & 1.0 \\
        $C_{\max}$  & 192   & $C_{\min}$ & 24 \\
        $D_{\text{port}}$ & 64 & $M$ & 256 \\
        \multicolumn{2}{c}{Maximum TCP Retransmissions} & \multicolumn{2}{c}{5} \\
        \multicolumn{2}{c}{Packet Length $L$} & \multicolumn{2}{c}{[768,1024]} \\
        \multicolumn{2}{c}{Protocols} & \multicolumn{2}{c}{TCP, UDP, SITP, SPAT} \\
        \multicolumn{2}{c}{Digital Modulation} & \multicolumn{2}{c}{16QAM} \\
        \hline
        \hline
    \end{tabular}
    \label{tab:spat_parameters}
\end{table}

\subsection{Latency Performance Analysis}

We compare the transmission latency of TCP, UDP, and SPAT. For each protocol, only a single packet is transmitted, while a large number of Monte Carlo trials are conducted to ensure statistical reliability. For TCP, both the three-way handshake and acknowledgement (ACK) procedures are taken into account; therefore, network fluctuations are incorporated into the latency evaluation. By contrast, UDP and SPAT perform direct transmission without feedback or retransmission. The effect of network jitter is explicitly considered, which is modeled by a truncated Gaussian distribution. The simulation parameters are summarized in Table \ref{latency}.

Fig.~\ref{fig:latency} presents the probability density functions (PDFs) and cumulative distribution functions (CDFs) of latency under a packet loss rate of 0.25. As shown in Fig.~\ref{fig:latency}(a), the latency distribution of TCP exhibits a pronounced long-tail characteristic, indicating substantial retransmission delays caused by repeated acknowledgement and timeout procedures. The corresponding CDFs in Fig.~\ref{fig:latency}(b) further support these observations. SPAT achieves significantly lower end-to-end latency than TCP, mainly because it avoids the handshake and ACK exchange procedures. Moreover, SPAT yields slightly lower latency than UDP, since it embeds explicit port-header information into the payload, thereby shortening the packet length.

\subsection{Reliability Performance Analysis}

We employ an image-oriented SemCom framework to validate the reliability advantages of the proposed SPAT system through end-to-end training. The semantic transceiver is developed based on the SwinJSCC architecture and further extended to accommodate digital communication scenarios. The simulation parameters are summarized in Table \ref{tab:spat_parameters}.

\begin{figure*}[t]
    \centering
    \begin{minipage}{0.30\textwidth}
        \centering
        \includegraphics[width=\textwidth]{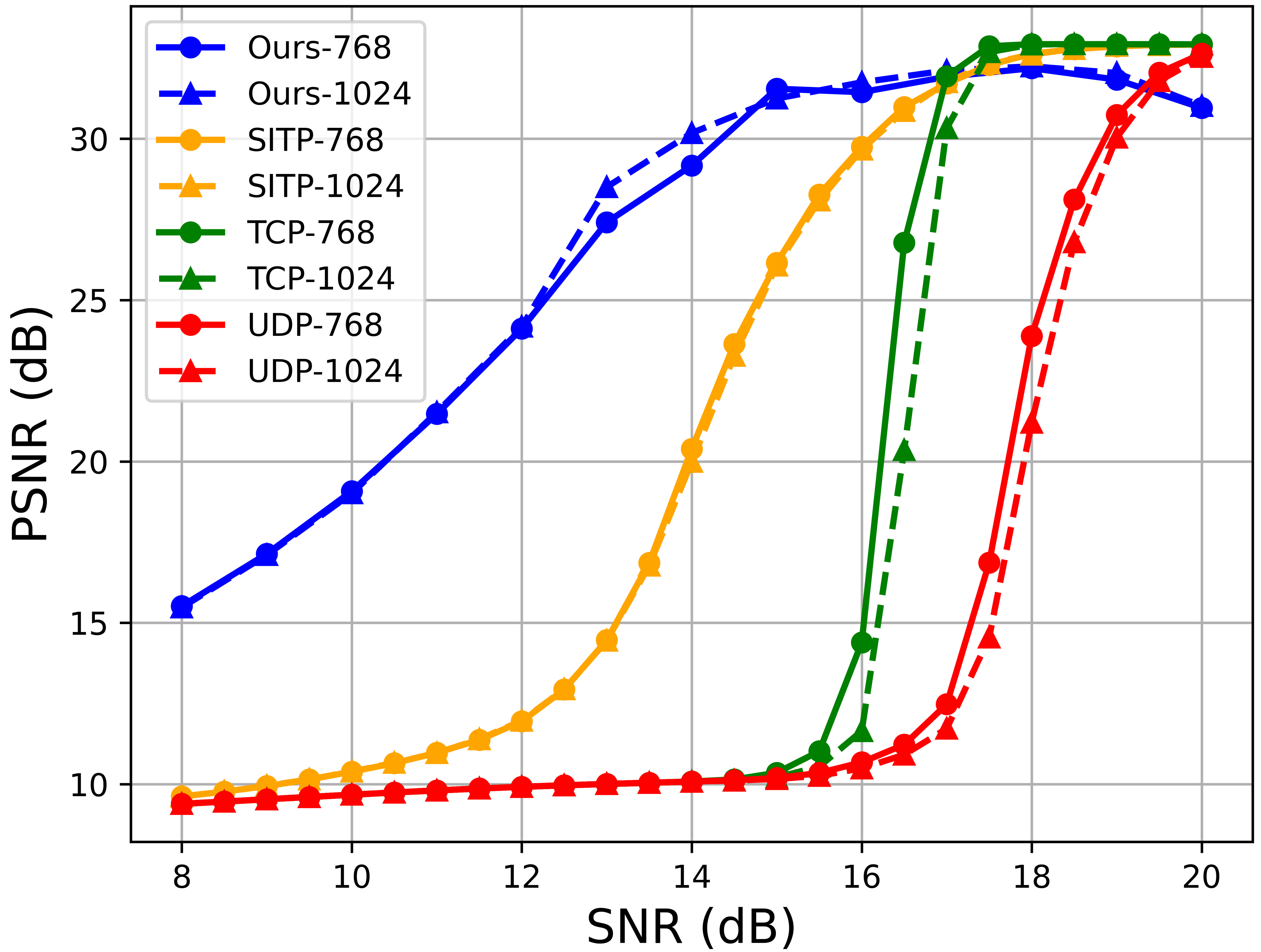}\\
        {\footnotesize (a) PSNR in the Uplink Scenario}
    \end{minipage}
    \hfill
    \begin{minipage}{0.30\textwidth}
        \centering
        \includegraphics[width=\textwidth]{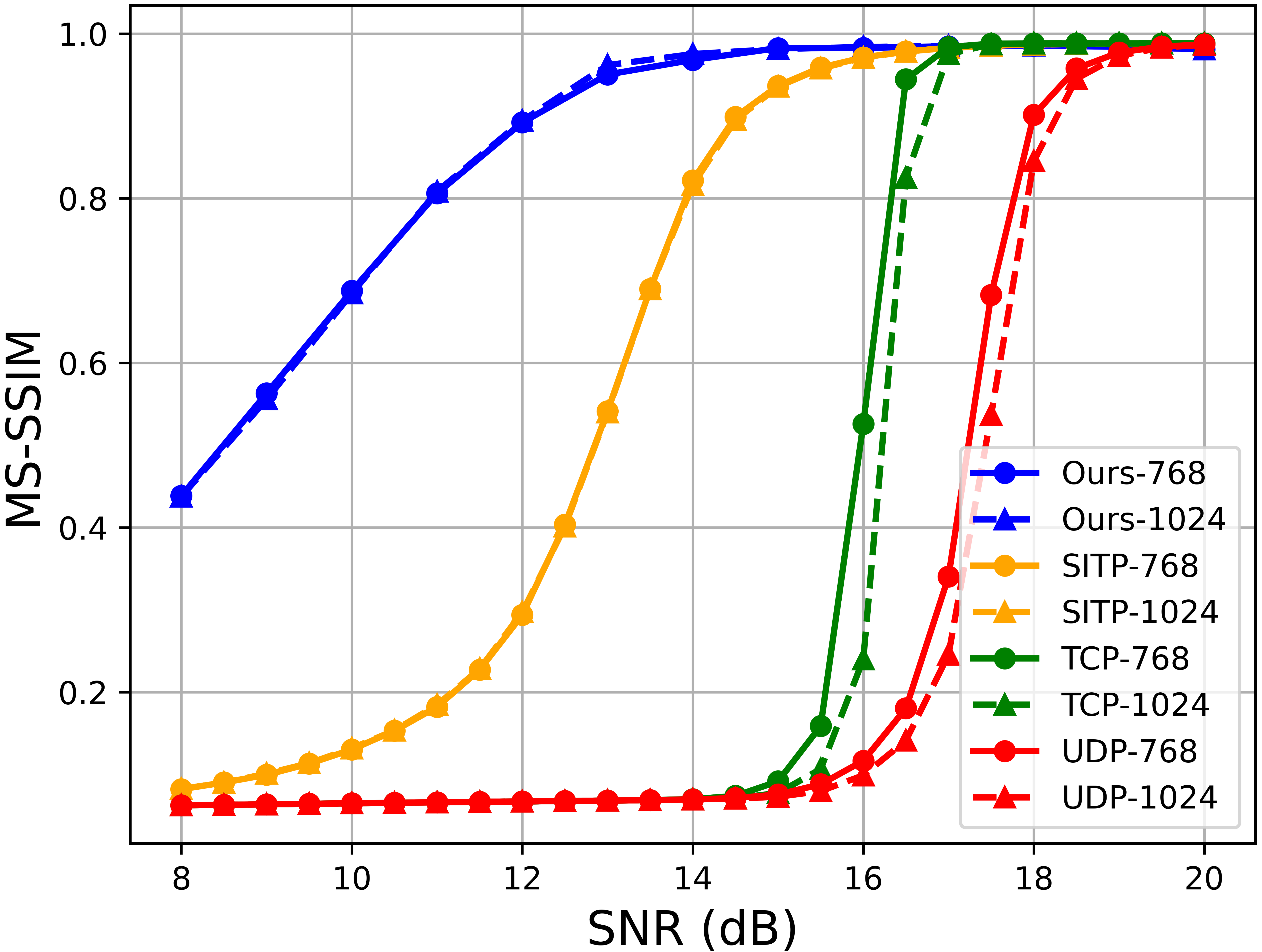}\\
        {\footnotesize (b) MS-SSIM in the Uplink Scenario}
    \end{minipage}
    \hfill
    \begin{minipage}{0.30\textwidth}
        \centering
        \includegraphics[width=\textwidth]{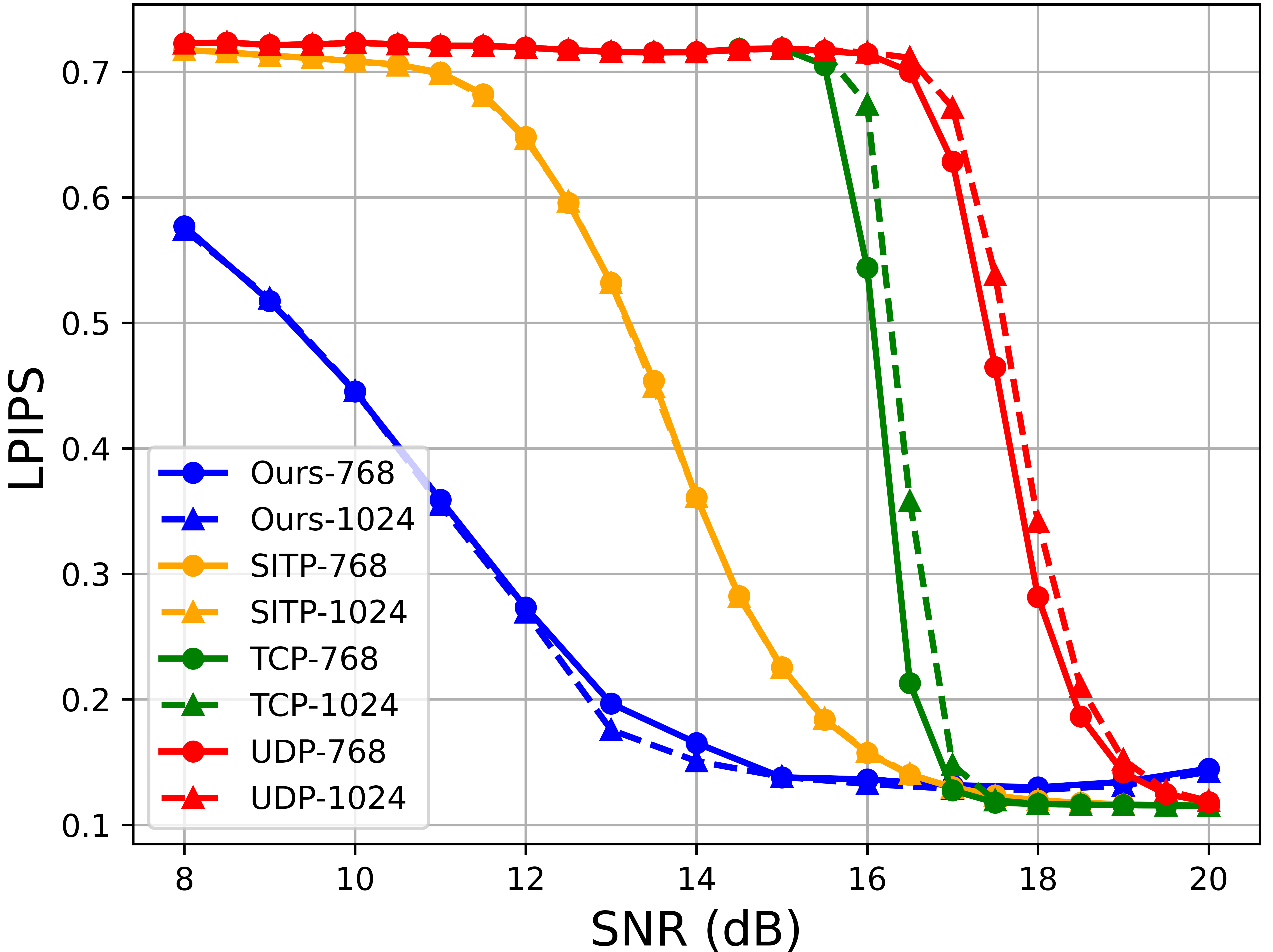}\\
        {\footnotesize (c) LPIPS in the Uplink Scenario}
    \end{minipage}

    \vspace{0.6em}

    \begin{minipage}{0.30\textwidth}
        \centering
        \includegraphics[width=\textwidth]{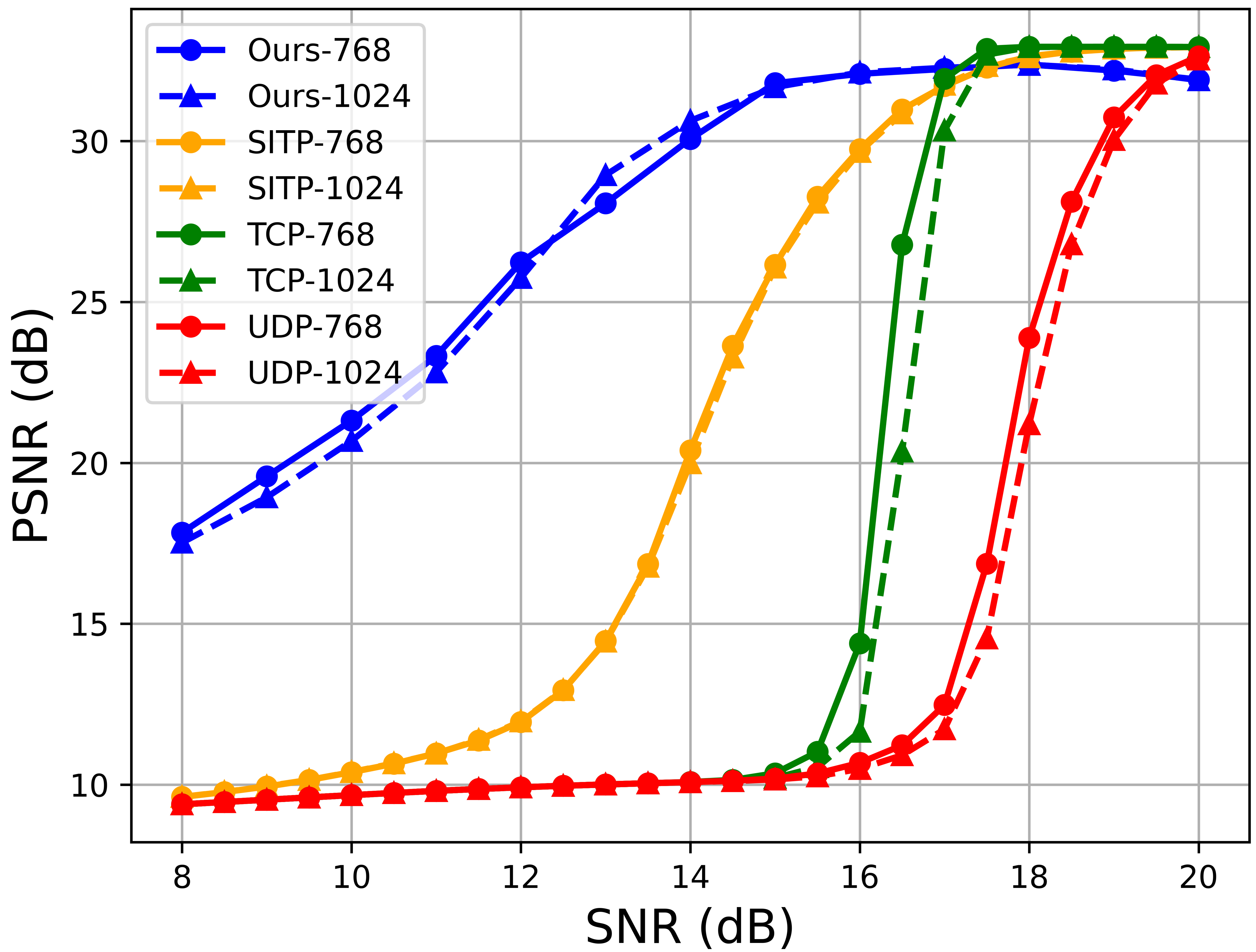}\\
        {\footnotesize (d) PSNR in the Downlink Scenario}
    \end{minipage}
    \hfill
    \begin{minipage}{0.30\textwidth}
        \centering
        \includegraphics[width=\textwidth]{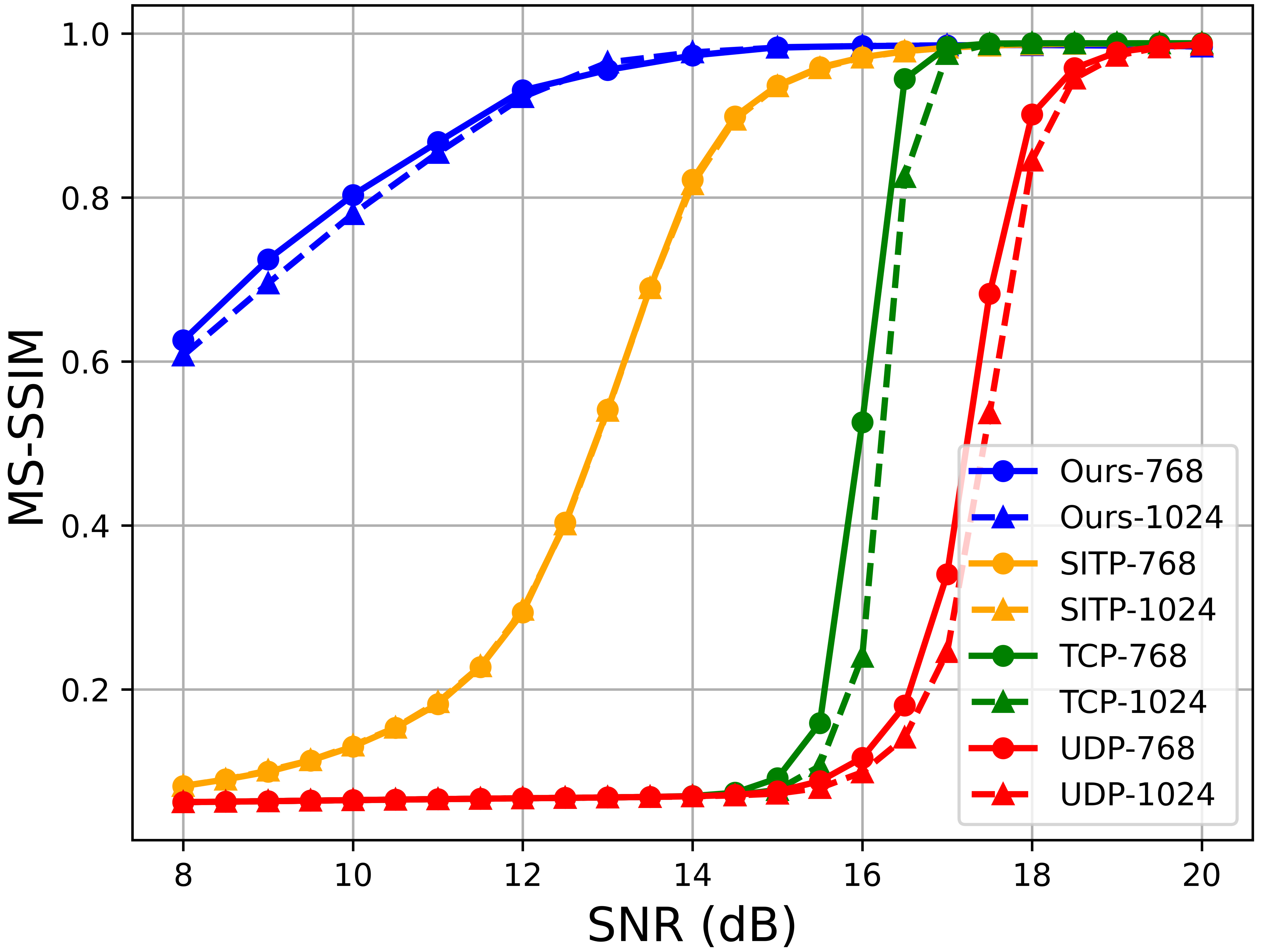}\\
        {\footnotesize (e) MS-SSIM in the Downlink Scenario}
    \end{minipage}
    \hfill
    \begin{minipage}{0.30\textwidth}
        \centering
        \includegraphics[width=\textwidth]{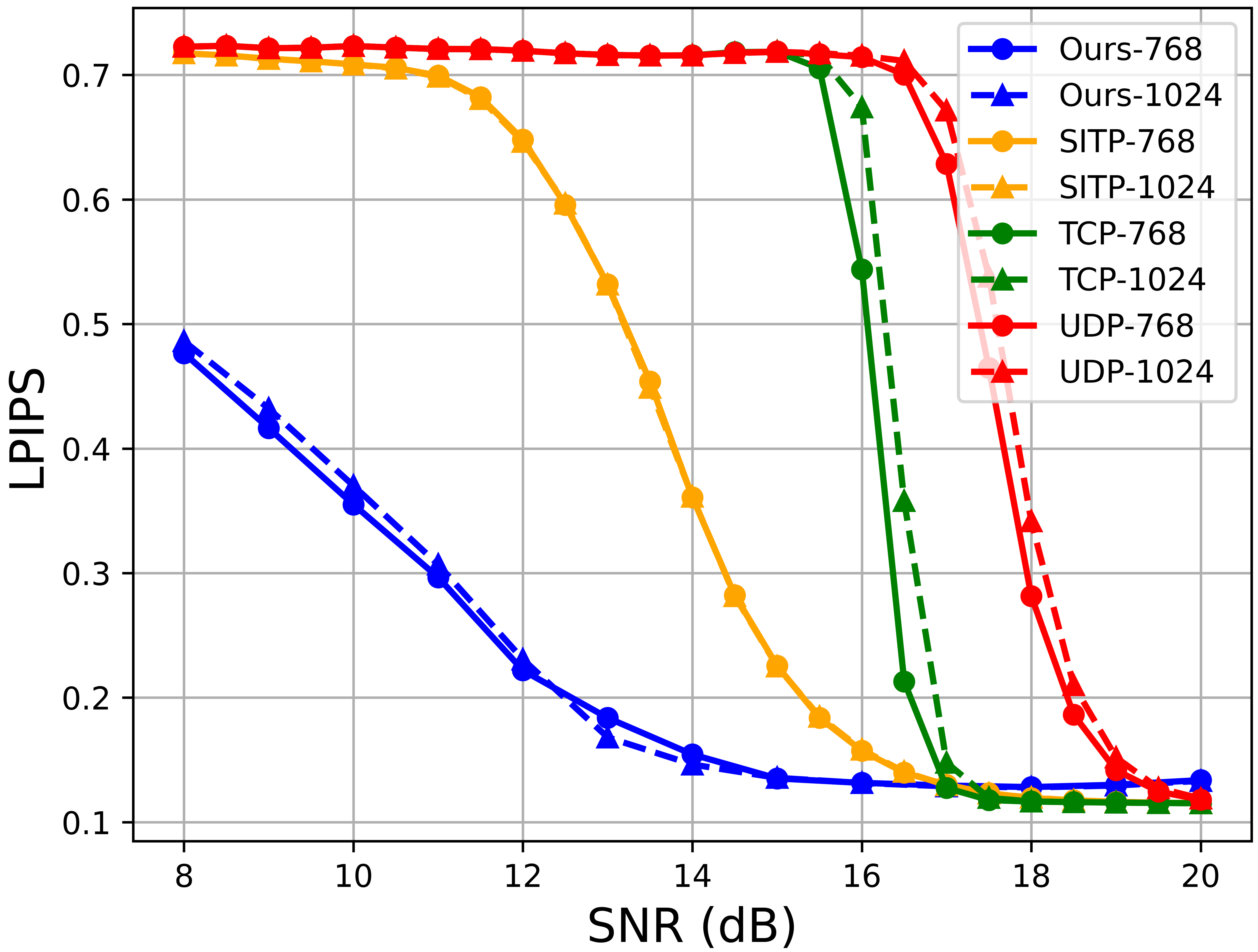}\\
        {\footnotesize (f) LPIPS in the Downlink Scenario}
    \end{minipage}

    \caption{Performance comparison among SPAT, SITP, TCP, and UDP over AWGN channels on the AFHQ dataset in both uplink and downlink scenarios. The results demonstrate that SPAT consistently outperforms the other baselines across different SNRs.}
    \label{fig:afhq_snr_performance}
\end{figure*}

\begin{figure*}[t]
    \centering
    \begin{minipage}{0.30\textwidth}
        \centering
        \includegraphics[width=\textwidth]{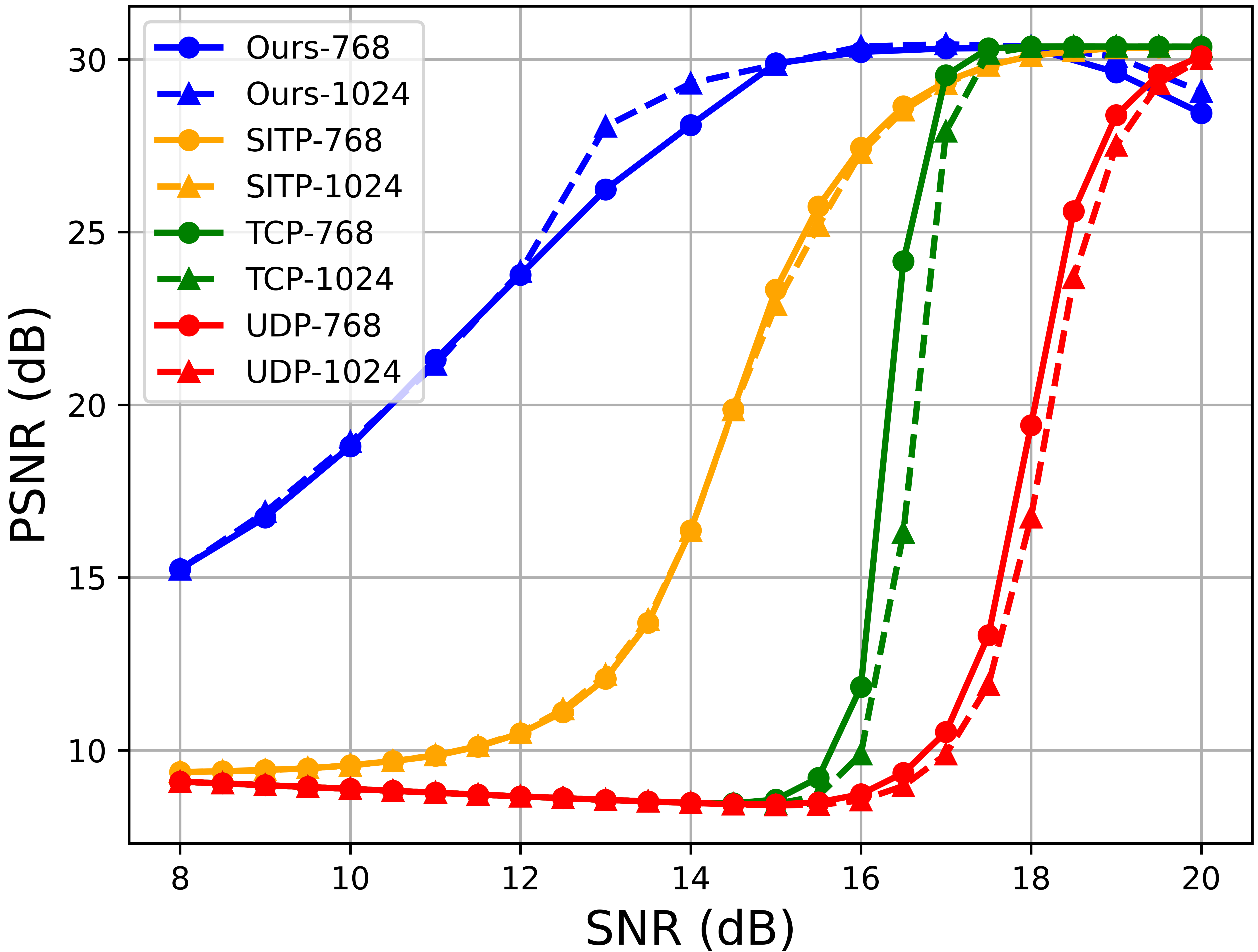}\\
        {\footnotesize (a) PSNR in the Uplink Scenario}
    \end{minipage}
    \hfill
    \begin{minipage}{0.30\textwidth}
        \centering
        \includegraphics[width=\textwidth]{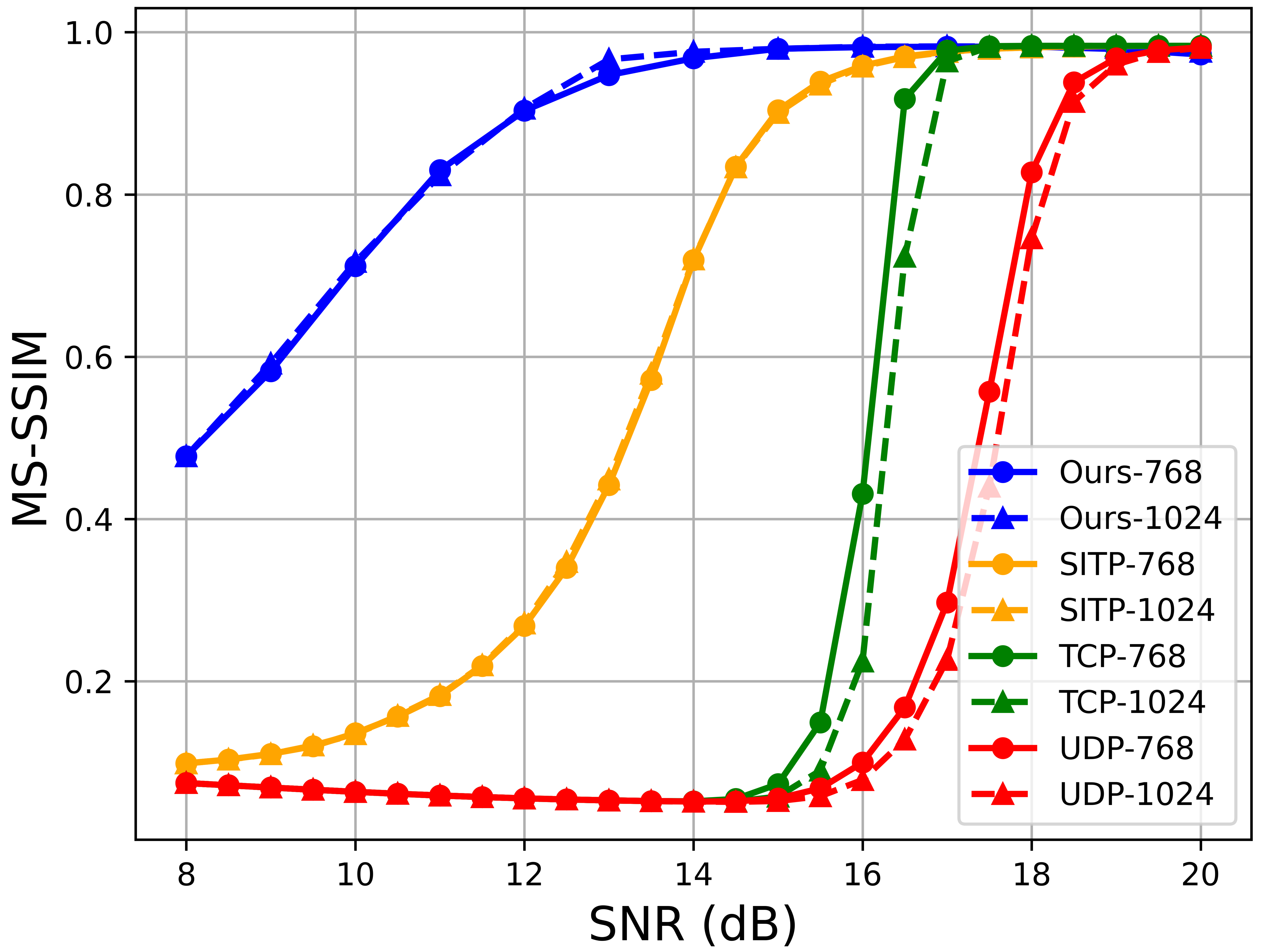}\\
        {\footnotesize (b) MS-SSIM in the Uplink Scenario}
    \end{minipage}
    \hfill
    \begin{minipage}{0.30\textwidth}
        \centering
        \includegraphics[width=\textwidth]{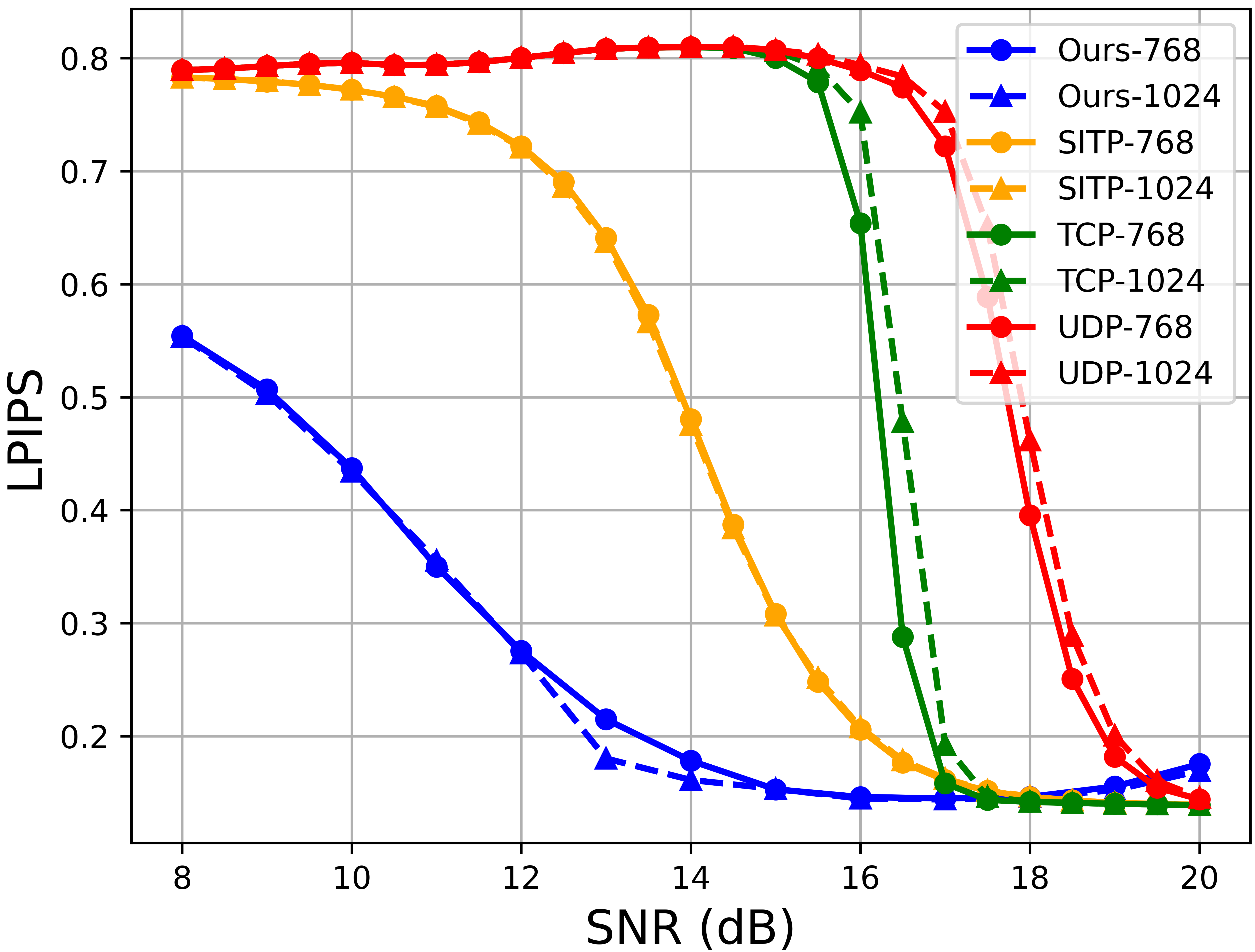}\\
        {\footnotesize (c) LPIPS in the Uplink Scenario}
    \end{minipage}

    \vspace{0.2em}

    \begin{minipage}{0.30\textwidth}
        \centering
        \includegraphics[width=\textwidth]{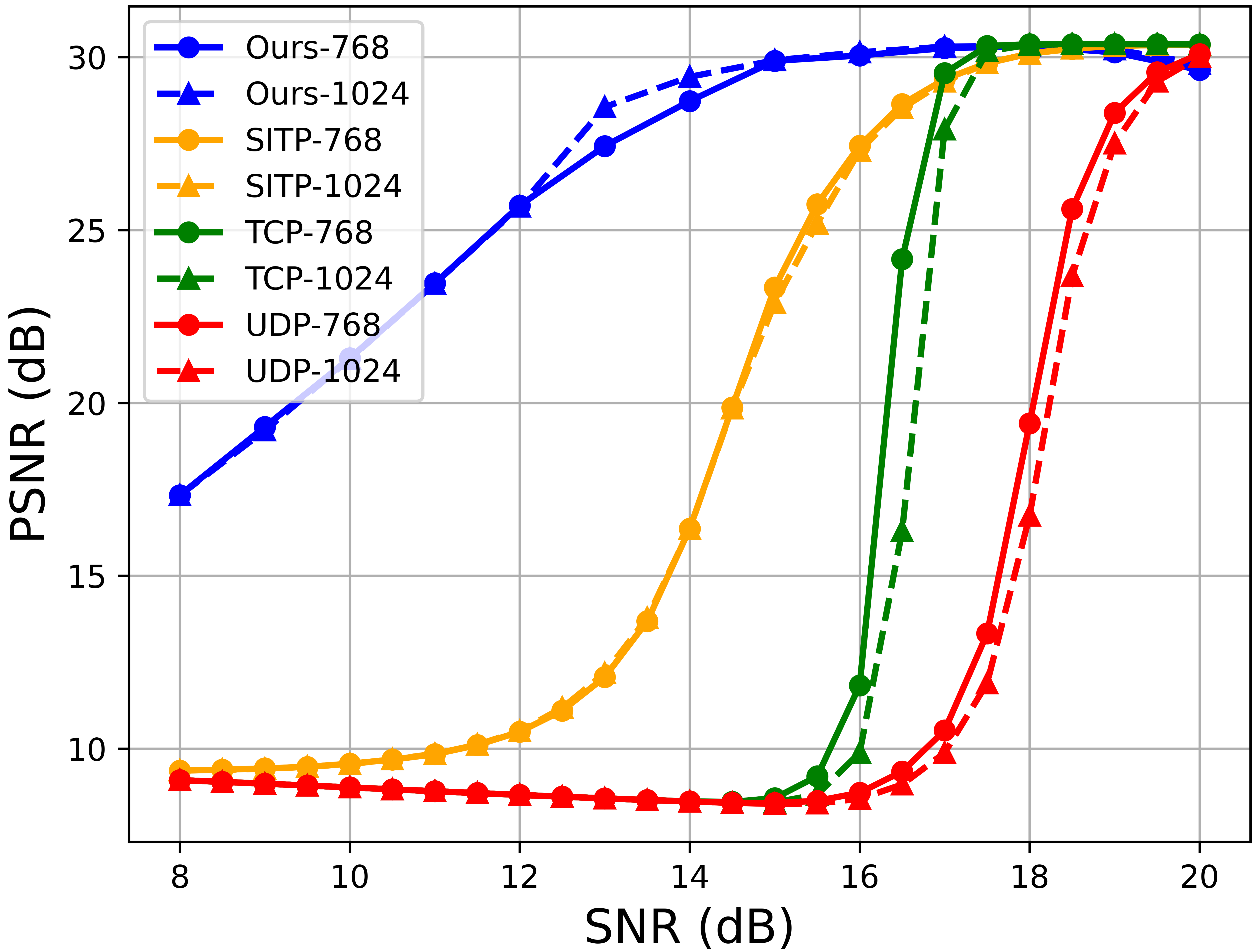}\\
        {\footnotesize (d) PSNR in the Downlink Scenario}
    \end{minipage}
    \hfill
    \begin{minipage}{0.30\textwidth}
        \centering
        \includegraphics[width=\textwidth]{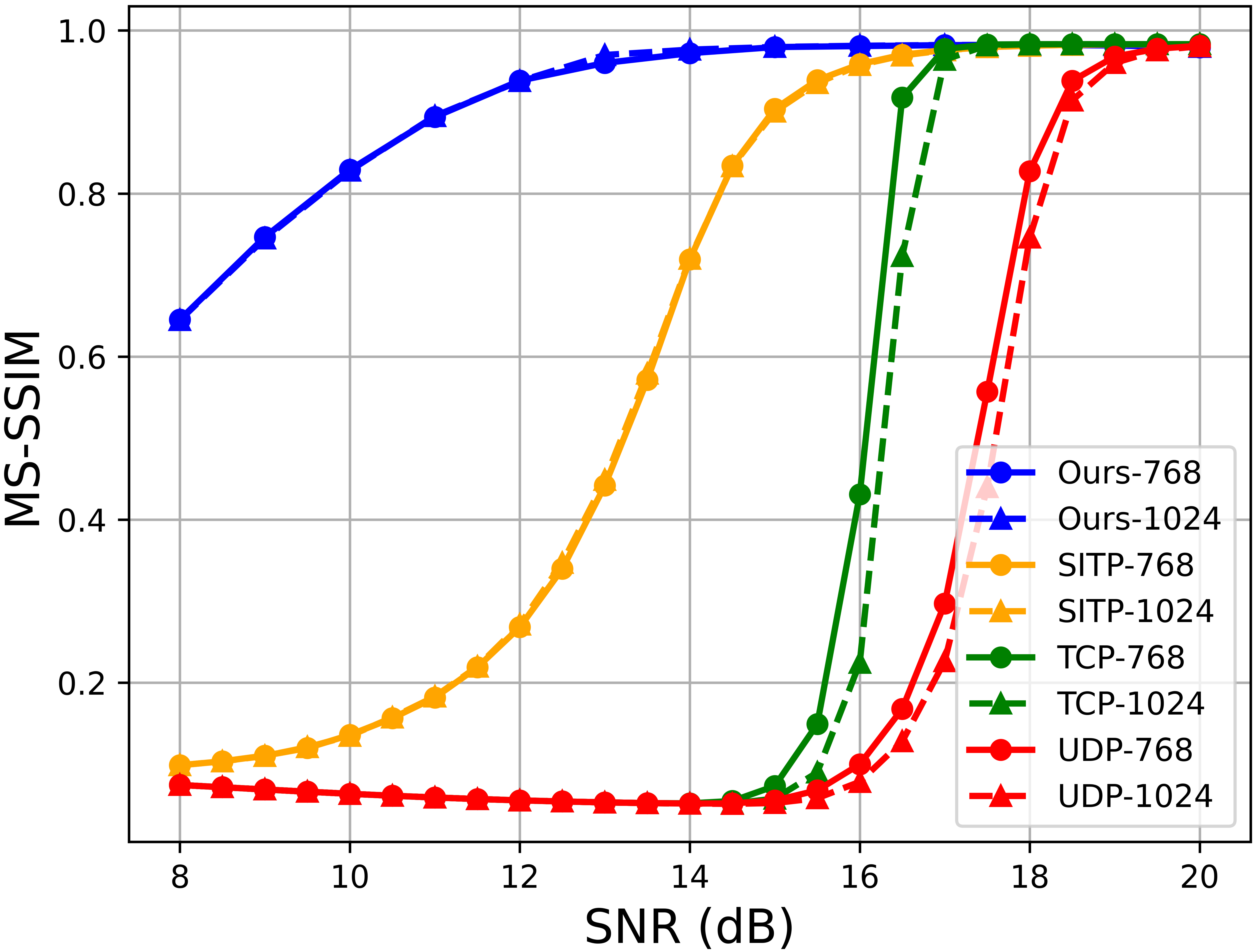}\\
        {\footnotesize (e) MS-SSIM in the Downlink Scenario}
    \end{minipage}
    \hfill
    \begin{minipage}{0.30\textwidth}
        \centering
        \includegraphics[width=\textwidth]{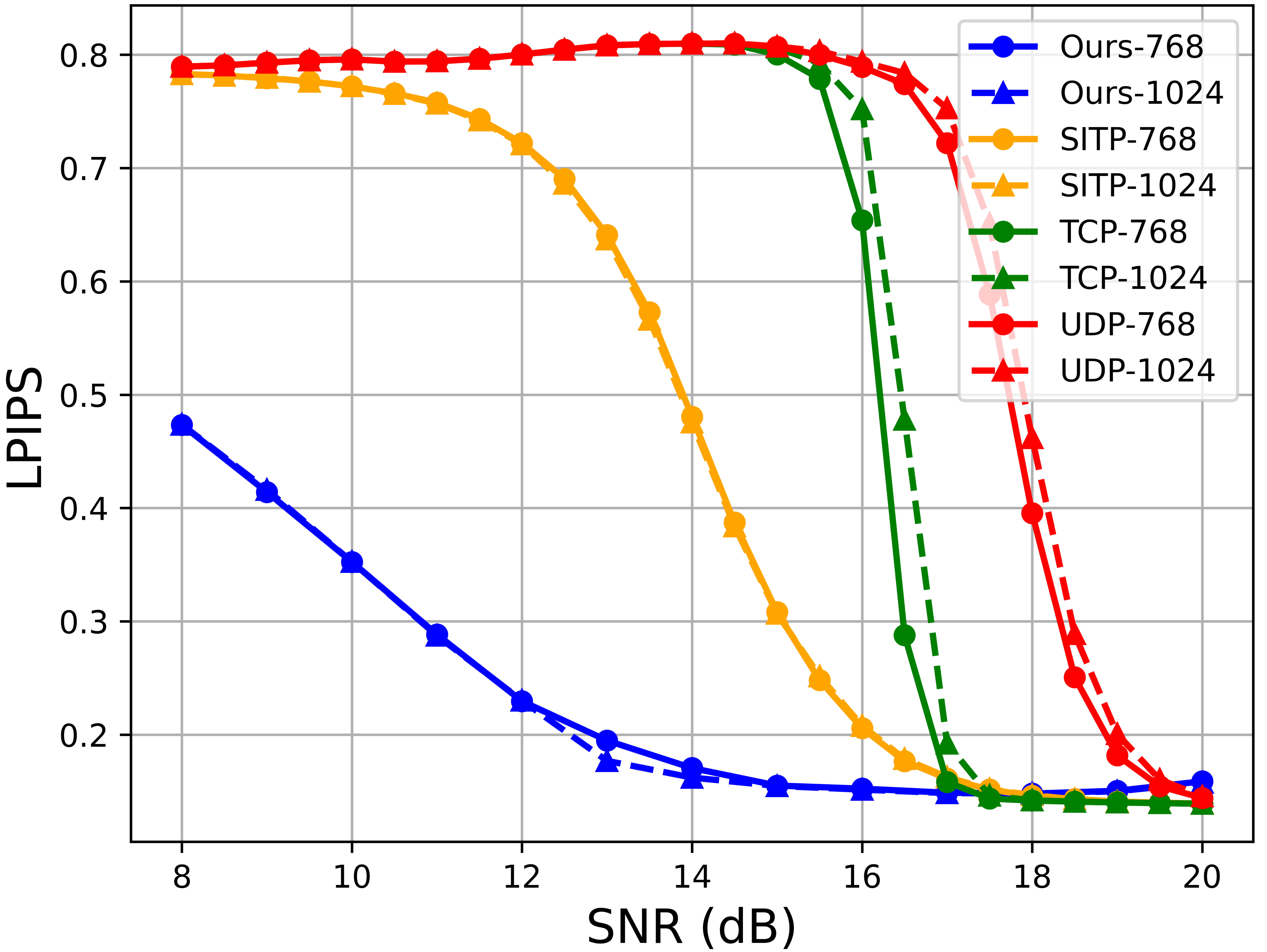}\\
        {\footnotesize (f) LPIPS in the Downlink Scenario}
    \end{minipage}

    \caption{Performance comparison among SPAT, SITP, TCP, and UDP over AWGN channels on the ImageNet10 dataset in both uplink and downlink scenarios. The results demonstrate that SPAT consistently outperforms the other baselines across different SNRs.}
    \label{fig:imagnet10_snr_performance}
\end{figure*}

As shown in Fig.~\ref{fig:afhq_snr_performance} and \ref{fig:imagnet10_snr_performance}, the proposed SPAT framework consistently achieves the best reconstruction performance across different SNRs. Specifically, SPAT attains higher PSNR and MS-SSIM values and lower LPIPS scores than SITP, TCP, and UDP, demonstrating superior robustness. This advantage mainly stems from the joint design of port-aware semantic transmission and adaptive-rate control, which preserves more informative semantic components under adverse channel conditions and thereby improves reconstruction reliability.

Compared with UDP and TCP, SPAT achieves clear performance gains across the entire SNR range. UDP directly discards corrupted packets, leading to severe quality degradation under noisy conditions, whereas SITP preserves corrupted semantic payloads for task-oriented reconstruction and thus enables graceful degradation. Building on SITP, SPAT further improves reliability through port-aware semantic embedding and adaptive-rate transmission, making it more robust to channel distortion and varying channel conditions. Furthermore, the superiority of SPAT is observed not only in the uplink but also in the downlink scenario, indicating that the proposed framework generalizes well across different communication directions. In the downlink, the destination-aware selective decoding mechanism further helps retain the desired semantic information at the target receiver, thereby supporting stable reconstruction quality without requiring explicit destination-port header information at the receiver.

\section{Real-World Experimental Results}



\begin{figure}[t]
    \centering
    \includegraphics[width=0.48\textwidth]{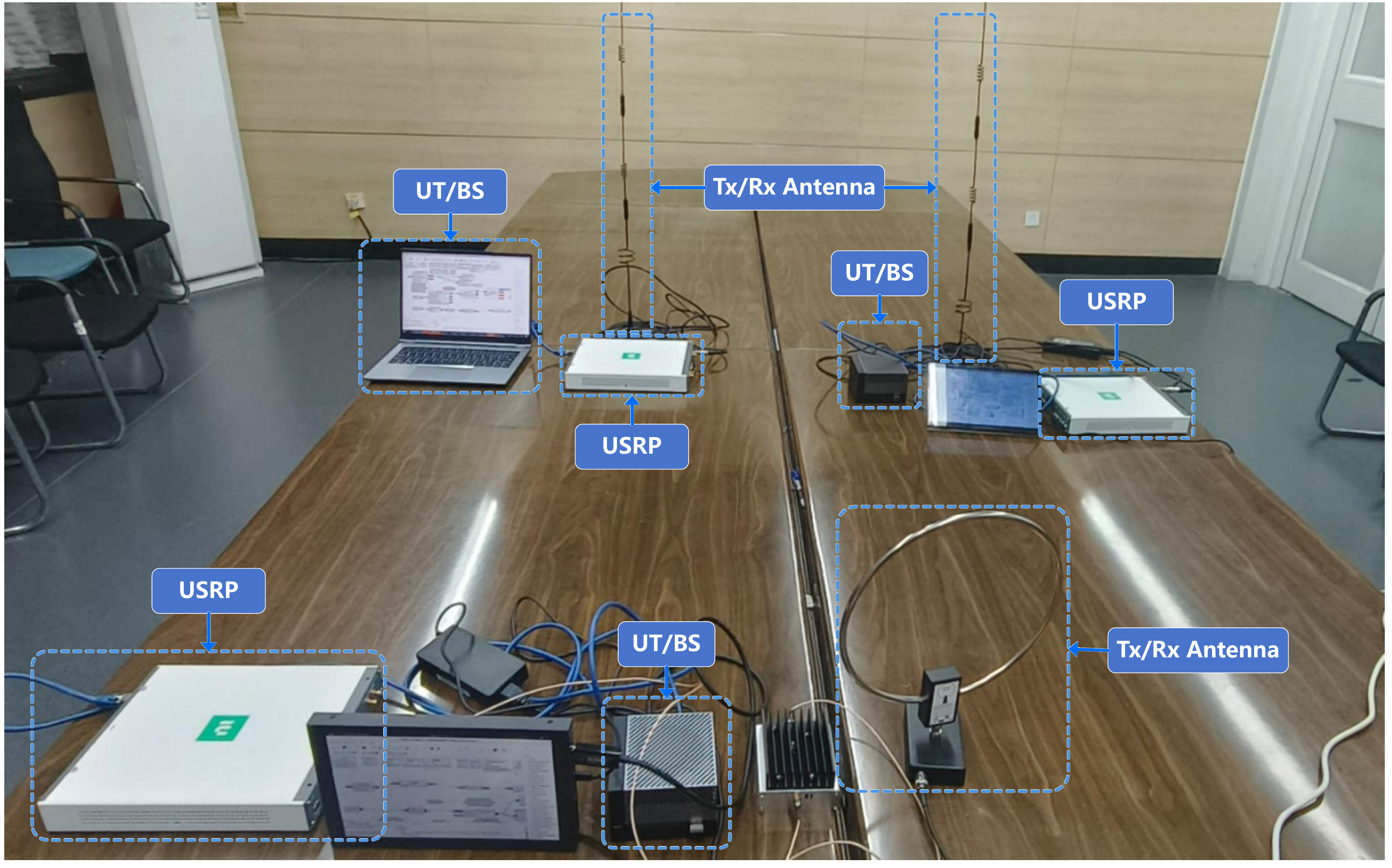}
    \caption{The real-world testbed for the proposed SPAT system.}
    \label{fig:realworld_setup}
\end{figure}

To further validate the practical effectiveness of the proposed SPAT framework, we implement a real-world SemCom experiment, as illustrated in Fig.~\ref{fig:realworld_setup}. The hardware platform consists of three USRPs, corresponding to two UTs and one BS, respectively, which is used to emulate both uplink and downlink transmission scenarios. The trained SemCom network is deployed on the NVIDIA Jetson AGX Orin platform, where semantic encoding, decoding, and quantization are executed at the device side. After quantization, the generated semantic bit streams are delivered to the USRP front-end for baseband processing. At the transmitter, the processed signals are further amplified by a power amplifier and radiated through the transmitting antenna over the wireless channel. At the receiver, the USRP performs analogue-to-digital conversion (ADC), synchronization, and demodulation, and then forwards the recovered bit streams to the semantic decoder.

Due to the instability of real-world wireless channels, packet loss rates (PLR) may fluctuate across repeated experiments. To ensure a fair and conservative comparison, we conduct multiple trials and report the SPAT results under relatively higher packet loss rates, while selecting the SITP and UDP results under relatively lower packet loss rates. This setting avoids overestimating the performance advantage of SPAT due to more favourable channel conditions.

\begin{figure}[t]
    \centering
    \begin{minipage}{0.5\textwidth}
        \centering
        \includegraphics[width=0.88\textwidth]{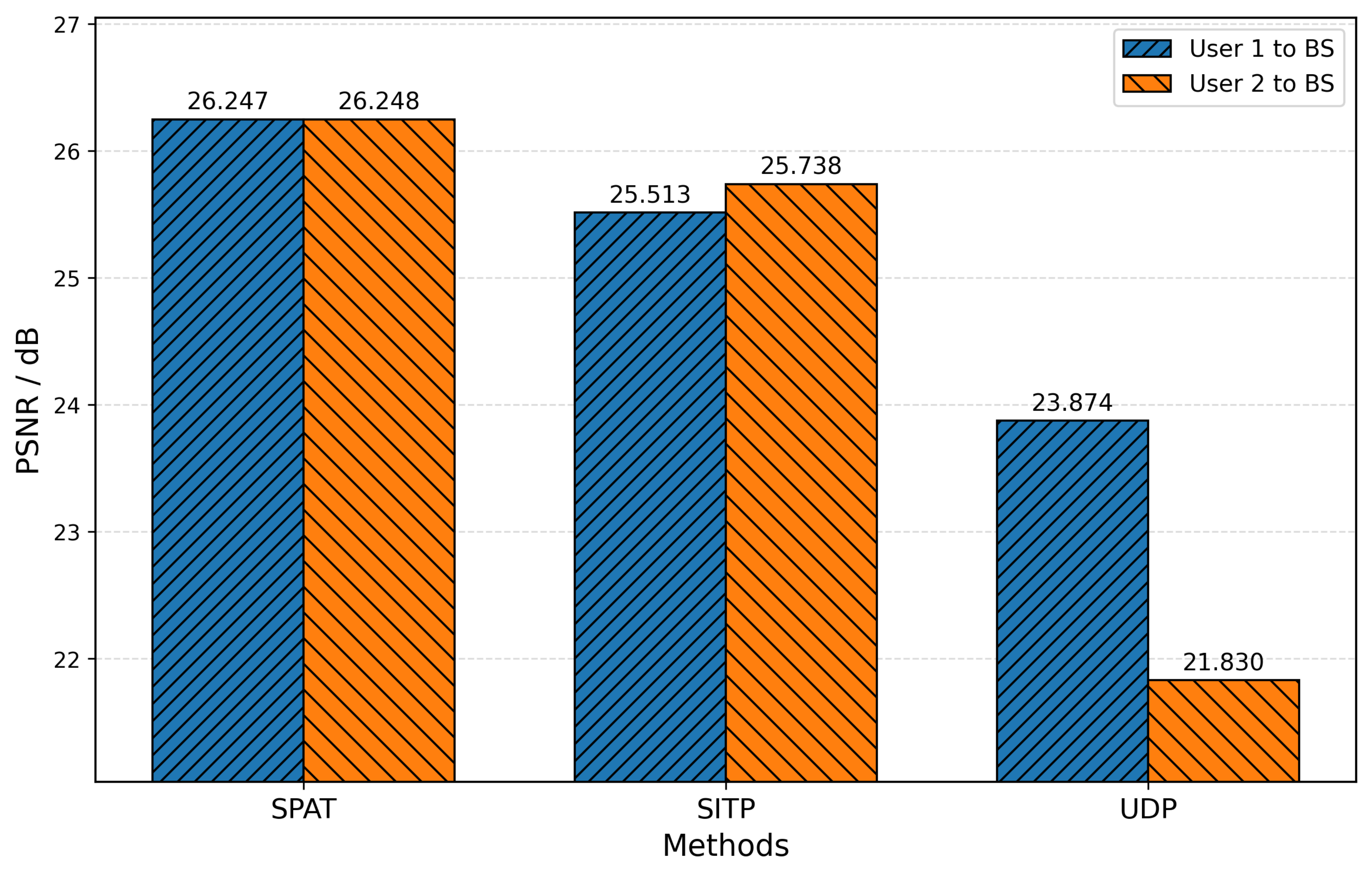} \\
        {\footnotesize (a) Real-world PSNR comparison in the uplink scenario.}
        \label{fig:realworld_setup1}
    \end{minipage}
    \begin{minipage}{0.5\textwidth}
        \centering
        \includegraphics[width=0.88\textwidth]{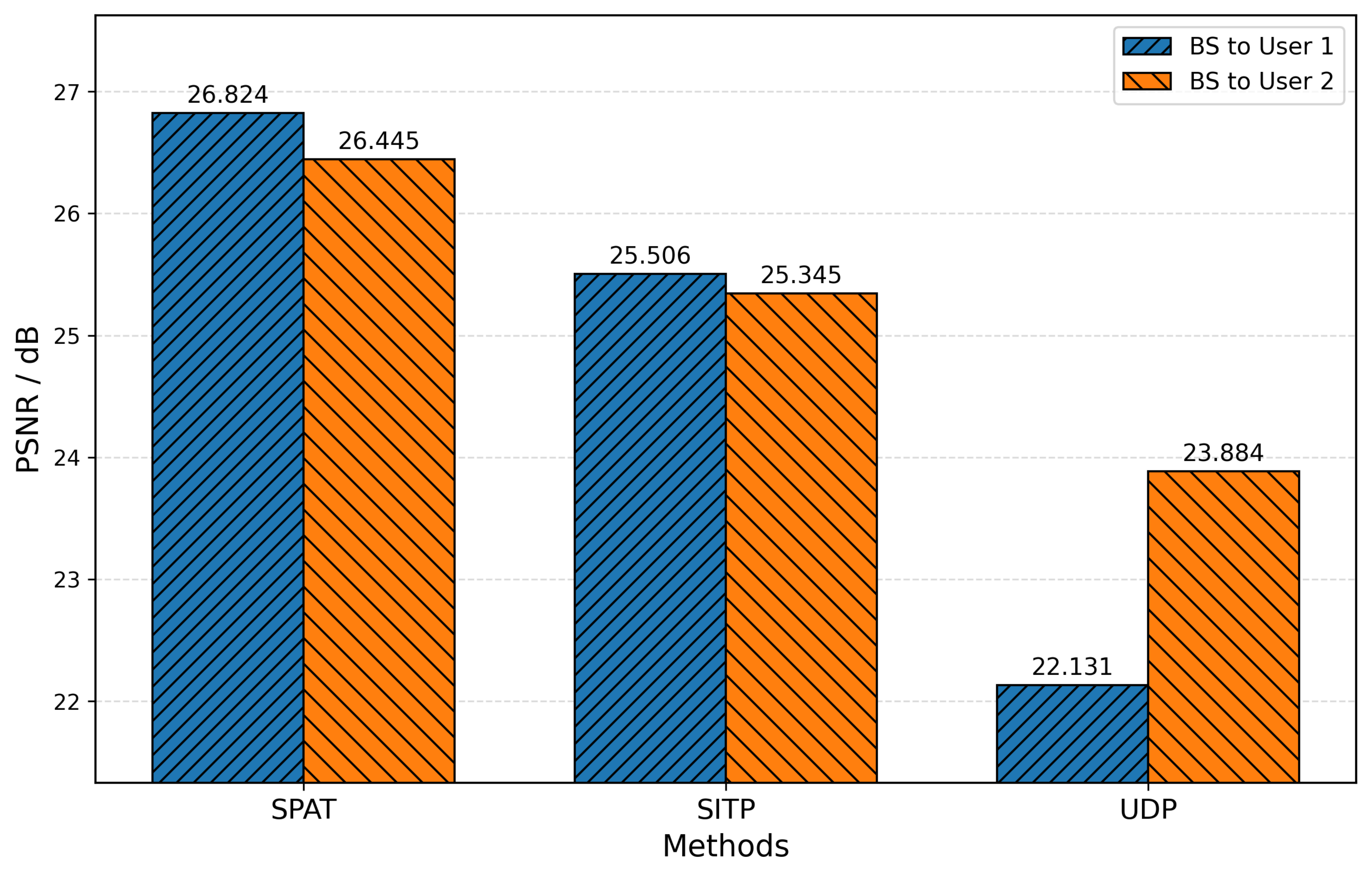} \\
        {\footnotesize (b) Real-world PSNR comparison in the downlink scenario.}
        \label{fig:realworld_setup2}
    \end{minipage}
    \caption{Real-world PSNR comparison of SPAT, SITP, and UDP in the uplink and downlink scenarios.}
    \label{fig:realworld_psnr}
\end{figure}

The real-world PSNR results for both uplink and downlink transmissions are shown in Table~\ref{tab:real_world_usrp_plr} and Fig.~\ref{fig:realworld_psnr}. In the uplink scenario, SPAT achieves an average PSNR of 26.786 dB, outperforming SITP and UDP. In the downlink scenario, SPAT attains an average PSNR of 26.870 dB, again exceeding the competing schemes. These results demonstrate that the proposed SPAT framework consistently delivers superior reconstruction quality under practical over-the-air transmission conditions, thereby confirming the effectiveness of its joint port-aware design and adaptive transmission mechanism in real-world deployments.


\captionsetup[table]{justification=centering, labelsep=space, textfont=sc}
\begin{table}[t]
    \renewcommand\arraystretch{1.7} 
    \setlength{\tabcolsep}{7.0pt}
    \caption{\\ Packet Loss Rates in Real-World USRP Experiments}
    \centering
    \begin{tabular}{ccccc}
        \hline
        \hline
        Scenario & Protocol & Link & PLR & PSNR / dB \\
        \hline

        \multirow{6}{*}{Uplink}

        & \multirow{2}{*}{SITP}
        & UT1 $\rightarrow$ BS & 2.943\% & 25.513 \\
        &
        & UT2 $\rightarrow$ BS & 2.824\% & 25.738 \\
        \cline{2-5}

        & \multirow{2}{*}{UDP}
        & UT1 $\rightarrow$ BS3 & 2.824\% & 23.874 \\
        &
        & UT2 $\rightarrow$ BS3 & 3.185\% & 21.830 \\
        \cline{2-5}

        & \multirow{2}{*}{\textbf{SPAT}}
        & \textbf{UT1} $\bm{\rightarrow}$ \textbf{BS} & \textbf{3.484\%} & \textbf{26.247} \\
        &
        & \textbf{UT2} $\bm{\rightarrow}$ \textbf{BS} & \textbf{3.628\%} & \textbf{26.248} \\
        \hline

        \multirow{6}{*}{Downlink}

        & \multirow{2}{*}{SITP}
        & BS $\rightarrow$ UT2 & 2.824\% & 25.806 \\
        &
        & BS $\rightarrow$ UT3 & 3.005\% & 25.345 \\
        \cline{2-5}

        & \multirow{2}{*}{UDP}
        & BS $\rightarrow$ UT2 & 2.943\% & 22.131 \\
        &
        & BS $\rightarrow$ UT3 & 2.704\% & 23.884 \\
        \cline{2-5}

        & \multirow{2}{*}{\textbf{SPAT}}
        & \textbf{BS} $\bm{\rightarrow}$ \textbf{UT2} & \textbf{3.125\%} & \textbf{26.824} \\
        &
        & \textbf{BS} $\bm{\rightarrow}$ \textbf{UT3} & \textbf{3.484\%} & \textbf{26.445} \\     
        \hline
        \hline
    \end{tabular}
    \label{tab:real_world_usrp_plr}
\end{table}

\section{Conclusion}

A novel transport-layer framework termed the Semantic Port-Aware Adaptive-Rate Transmission Protocol (SPAT) is proposed to improve the reliability and efficiency of SemCom systems. Unlike conventional transport mechanisms that rely on explicit port headers and bit-level validation, the proposed framework jointly embeds source and destination port information into semantic representations, thereby reducing header overhead and enhancing robustness to channel distortion. Furthermore, differentiated semantic processing mechanisms are developed for uplink and downlink scenarios. In addition, an adaptive-rate control mechanism is incorporated to dynamically adjust the number of transmitted semantic channels according to channel conditions and feature importance. Experimental results demonstrate that SPAT consistently outperforms TCP, UDP, and SITP in reconstruction quality across different SNRs while maintaining low-latency transmission.

\end{document}